\documentclass[a4,journal]{IEEEtran}
\usepackage{amsmath,amssymb,dsfont,stfloats,color,url,mathtools}
\usepackage[]{graphicx}
\usepackage{comment}
\usepackage{cuted}
\usepackage{amsmath}
\usepackage{subcaption} 
\usepackage{enumerate}
\usepackage{bbm}
\usepackage{wasysym,esint}
\usepackage{cite}
\usepackage{caption}
\usepackage{subcaption}
\usepackage{gensymb} 
\usepackage[boxruled,linesnumbered]{algorithm2e}
\usepackage{algorithm2e}
\usepackage{lipsum}

\DeclareGraphicsExtensions{.eps,.pdf,.png,.jpg,.gif,.jpeg,.pstex}

\newtheorem{rem}{Remark}

\DeclareMathOperator*{\argmax}{arg\,max}

\newcommand{\twodots}{\mathinner {\ldotp \ldotp}}

\newfont{\bbb}{msbm10 scaled 700}

\newfont{\bb}{msbm10 scaled 1100}
\newcommand{\CC}{\mbox{\bb C}}

\newcommand{\RR}{\mbox{\bb R}}

\newcommand{\av}{{\bf a}}
\newcommand{\bv}{{\bf b}}

\newcommand{\ev}{{\bf e}}

\newcommand{\gv}{{\bf g}}
\newcommand{\hv}{{\bf h}}

\newcommand{\jv}{{\bf j}}

\newcommand{\mv}{{\bf m}}
\newcommand{\nv}{{\bf n}}

\newcommand{\pv}{{\bf p}}
\newcommand{\qv}{{\bf q}}
\newcommand{\rv}{{\bf r}}
\newcommand{\sv}{{\bf s}}

\newcommand{\uv}{{\bf u}}
\newcommand{\wv}{{\bf w}}
\newcommand{\vv}{{\bf v}}
\newcommand{\xv}{{\bf x}}

\newcommand{\zerov}{{\bf 0}}

\newcommand{\Am}{{\bf A}}

\newcommand{\Gm}{{\bf G}}
\newcommand{\Hm}{{\bf H}}
\newcommand{\Id}{{\bf I}}

\newcommand{\Pm}{{\bf P}}

\newcommand{\Tm}{{\bf T}}
\newcommand{\Um}{{\bf U}}
\newcommand{\Wm}{{\bf W}}

\newcommand{\Ac}{{\cal A}}

\newcommand{\muv}{\hbox{\boldmath$\mu$}}

\newcommand{\Omegam}{\hbox{\boldmath$\Omega$}}
\newcommand{\Xim}{\hbox{\boldmath$\Xi$}}

\newcommand{\diag}{{\hbox{diag}}}
\renewcommand{\det}{{\hbox{det}}}

\renewcommand{\Re}{{\rm Re}}

\newcommand{\herm}{{\sf H}}

\newcommand{\transp}{{\sf T}}
\renewcommand{\vec}{{\rm vec}}

\newcommand{\SNR}{{\sf SNR}}

\usepackage{stmaryrd} 

\usepackage{hyperref}
\hypersetup{
    bookmarks=true,         
    unicode=false,          
    pdftoolbar=true,        
    pdfmenubar=true,        
    pdffitwindow=false,     
    pdfstartview={FitH},    
    pdfnewwindow=true,      
    colorlinks=true,       
    linkcolor=red,          
    citecolor=green,        
    filecolor=blue,      
    urlcolor=blue           
}

\title{
Codebook Design and Baseband Precoding for  
Pragmatic Array-Fed RIS Hybrid Multiuser MIMO} 

\author{\IEEEauthorblockN{Krishan Kumar Tiwari\IEEEauthorrefmark{1}, \emph{Senior Member, IEEE}, and Giuseppe Caire\IEEEauthorrefmark{1}, \emph{Fellow, IEEE}} \thanks{\IEEEauthorrefmark {1}Technical University of Berlin, Germany. Email ids: lastname@tu-berlin.de. 
The work was supported by BMBF Germany in the program of ``Souverän. Digital. Vernetzt.'' Joint Project 6G-RIC (Project ID 16KISK030). \emph{Part of this work was
presented at the 101st IEEE Vehicular Technology Conference (VTC2025-Spring) \cite{VTC2025}.}
}}

\begin{document}

\bstctlcite{BSTcontrol} 

\maketitle

\begin{abstract}
In our previous work \cite{NewOldIdea}, we introduced a hardware-efficient and power-efficient pragmatic architecture for hybrid digital-analog (HDA) multiuser MIMO (MU-MIMO) based on the stacking of identical basic modules, each formed by a small active multi-antenna feeder (AMAF) placed in the near field of much larger reflective intelligent surface (RIS). Each AMAF is driven by one independent RF chain and conveys one 
spatial data stream. By stacking $K$ such modules at the BS, a spatial multiplexing gain of $K$ can be achieved. The work in \cite{NewOldIdea} focused on the design of the basic module and on the analysis of its power and hardware efficiency in comparison to a classical active array counterpart. However, the system performance in terms of MU-MIMO spectral efficiency was evaluated in a simple scenario of pure line-of-sight (LOS) propagation. 

In this work, we extend our approach in several ways. First, we
propose a novel simple and pragmatic method for designing phase-only flat-top beams 
for the basic AMAF-RIS module, enabling wide angular coverage with small passband ripple and low sidelobes. Our design allows the implementation of hierarchical beamforming codebooks for efficient 
beam acquisition. Then, we consider MU-MIMO system performance with realistic mmWave multipath channels including both a LOS component and non-LOS (NLOS) components, following a 3D von Mises-Fisher (vMF) distribution. We propose a pragmatic and low-complexity approach to HDA MU-MIMO that includes user-beam association via a standard beam acquisition phase, dynamic selection of MU-MIMO user groups by choosing one user per beam, estimation of the resulting {\em effective} baseband MIMO channel using standard 3GPP pilots, and MU-MIMO downlink transmission using zero-forcing precoding with 
per-antenna port power constraint applied in the baseband. Our results demonstrate that the proposed architecture achieves high spectral efficiency and spatial multiplexing gain while retaining the low hardware complexity and high power efficiency of our basic pragmatic approach. Importantly, the proposed approach is fully compliant with the basic mechanisms of beam acquisition and sounding reference signaling of the 3GPP 5GNR standard.
\end{abstract}

\begin{IEEEkeywords}
Reflectarrays, array-fed RIS, multiuser MIMO, hybrid digital-analog precoding, hierarchical beamforming, flat-top beams, von Mises-Fisher (vMF) distribution.
\end{IEEEkeywords}

\section{Introduction}  
\label{sec:intro}

\IEEEPARstart{T}{he} relentless demand for higher data rates and lower latency in wireless networks \cite{ofcom_connected_nations_2025} is driving the exploitation of centimeter waves, millimeter waves, 
and sub-terahertz frequencies. At these bands, achieving a sufficient link budget necessitates the use of electrically large antenna arrays to provide high beamforming gain. Our recent work \cite{NewOldIdea} introduced a novel, power- and hardware-efficient modular architecture based on stacking $K\geq 1$
basic modules, where each module is formed by a small active multi-antenna feeder (AMAF) placed 
in the near-field of a larger passive RIS in order to form a highly directional, steerable beam. 
Each AMAF-RIS module is driven by an individual baseband antenna port, in a one-stream-per-subarray (OSPS) configuration.  In \cite{NewOldIdea} we demonstrated excellent performance of multiuser MIMO (MU-MIMO) in pure LOS conditions by simply scheduling multiuser groups of $K$ users with sufficient angular separation, where no further baseband precoding was necessary. 

While the performance in LOS is compelling, practical cellular deployments, even at high frequencies, are characterized by multipath propagation where scattering and reflections create non-line-of-sight (NLOS) propagation paths. This fundamentally alters the multiuser interference landscape: unlike in the LOS-only case, multipath propagation causes multiuser interference even in the presence of users with large angular separation of their LOS components. Consequently, the pure RF beamforming approach that sufficed in \cite{NewOldIdea} becomes inadequate, necessitating more advanced digital baseband precoding in a 
hybrid digital-analog (HDA) scheme. 

To this purpose, we propose a pragmatic two-step approach. Consider a cell sector with $U \gg K$ users. Using a standard standard beam acquisition procedure as specified in 5GNR \cite{Book5GNR,SRSref, TR38802}, the $U$ users are associated to the pre-designed beams of a beamforming codebook. Since the position of the users in the sector evolves on a time scale of the order of tens of 
seconds, this procedure can be repeated periodically at low duty cycle in order to maintain the user-beam assignment updated. Depending on traffic demands, and operating on a much faster time scale of the order of a few milliseconds, a scheduler selects dynamically multiuser groups of $K$ users, such that the selected users belong to different beams. For the given selected multiuser group, the concatenation of the AMAF-RIS beamforming and the physical channel forms a $K \times K$ baseband frequency-selective 
{\em effective channel} that can be estimated by standard SRS piloting  \cite{Book5GNR,SRSref, TR38802}. The baseband digital precoder operates on such estimated effective channel. 

We notice that the high-gain, "pointy" principal eigenmode (PEM) beams optimized in \cite{NewOldIdea} for data transmission are ill-suited for beam acquisition and sector-level coverage, because their 
very ``pointy'' beamwidth would require an impractically large number of training beams to cover the entire cell sector, leading to prohibitive acquisition latency and overhead. Hence, the first contribution of this present work is to develop a methodology to design a hierarchical beamforming codebook enabling efficient beam acquisition by bisection methods (e.g., see \cite{Hybrid_mmWave}).

Designing such beams is a non-trivial, high-dimensional, non-convex problem, exacerbated by the fixed, non-uniform amplitude taper imposed by the PEM design on the RIS (see \cite{NewOldIdea}). We circumvent this complexity by a novel pragmatic method for linear arrays, combining binary phase grouping, a parametric phase perturbation function for beam widening, and a final local optimization step. This approach yields 
beams with a suitable width and a uniform main lobe gain, steep transitions, and low sidelobes.  

Our second contribution is the integration of a realistic and geometrically consistent 3D multipath channel model. To move beyond simplistic isotropic scattering models, we generate the directions of the scatterers using the von Mises-Fisher (vMF) distribution on the sphere \cite{MarzettaVMF, JohnVMF}. This model allows us to simulate realistic, non-isotropic clusters of scatterers, where the concentration parameter can be linked to the scatterers' physical size and distance from the base station (BS). This provides a rigorous foundation for evaluating system performance in a controlled yet representative multipath setting. Notice that for the system performance evaluation, it is {\em essential} to have geometric consistency between the geographically distributed users, i.e., all users in the sector are affected (up to pathloss and blockage) by a common set of scatterers, whose directions seen from the reference system of the BS array follows the vMF distribution.  

Our third contribution is a comprehensive system-level performance evaluation that incorporates user-beam association, effective baseband channel estimation, and hybrid digital analog (HDA) precoding. We demonstrate a complete system operation including:
1) A standard beam acquisition phase that allows each user to estimate the channel quality corresponding to each beam in the codebook (repeated once per user placement round); 
2) In each downlink slot, the BS schedules a MU-MIMO user group, choosing one user per beam; 
3) The corresponding frequency-selective {\em effective} baseband channel matrix formed by the concatenation of the physical multipath propagation channel from the RIS elements to the users' antennas, and the AMAF-RIS beamforming configuration for the selected beams is estimated via standard SRS uplink pilots and channel reciprocity, according to the well-known 5GNR 3GPP protocol \cite{TR38802}; 
4) Based on the acquired CSI, a Zero-Forcing (ZF) precoder with per-antenna port power constraint is applied in the digital baseband domain in order to eliminate the multiuser interference between the downlink data streams.

Our numerical results confirm that the proposed enhancements make the AMAF-RIS architecture a versatile and powerful solution.  Crucially, we demonstrate that while RF beamforming alone fails in multipath, the concatenation of analog beam selection and baseband ZF precoding successfully restores high per-user spectral efficiency. 

To further stress the relevance of the proposed approach, notice that our scheme does not use any exceedingly computationally complex alternating optimization of the analog (RIS coefficients) and digital (baseband precoding) coefficients, which is routinely proposed in works on HDA MU-MIMO \cite{ImPractical_Hybrid_Opt6, ImPractical_Hybrid_Opt5, ImPractical_Hybrid_Opt3, ImPractical_Hybrid_Opt2}.
In addition, many of these works assume that the high-dimensional channel between the analog array elements and the user antennas can be somehow estimated, which is in fact a very hard and practically infeasible task in the short time and low SRS pilot overhead regime of 3GPP. In contrast, our approach requires only the estimation of the low-dimensional effective channel, which is fully compatible with today's systems.

The paper is organized as follows: Section \ref{sec:system_model} summarizes the system model, Section \ref{sec:flat_top_codebooks} presents the  pragmatic flat-top beamforming and codebook design. Section \ref{sec:system} details the MU-MIMO system set-up. Section \ref{sec:results} presents numerical results, and Section \ref{sec:CONC} concludes the work.

{\bf Mathematical Notations:} $\text{x}^*$ denotes complex conjugate, $\|\xv\|$ is the Euclidean norm of a vector $\xv$,  $\odot$ denotes Hadamard (elementwise) product, $[\cdot]^\transp$ is transpose, $[\cdot]^\herm$ is the Hermitian transpose, $|\xv|$ and $\angle \xv$ are vectors/matrices containing the magnitudes and the angles (phases) of the elements in a complex-valued vector/matrix $\xv$, respectively. $\mathbbm{1}\{\Ac\}$ denotes the indicator function for a condition $\Ac$, equal to 1 if the condition is true and 0 if it is false. For two vectors $\pv, \qv$, 
$\pv \cdot \qv$ denotes the inner product in the appropriate (real or complex) Hilbert space, and $\pv \times \qv$ denotes the Cartesian product, represented by an array with $(i,j)$ elements $p_i q_j$.  We use $\uv = \vec(\Um)$ to indicate the formatting of a matrix $\Um$ by stacking its columns on top of each other, and $\Um = \vec^{-1}(\uv)$ the inverse formatting, where the dimensions are clear from the context. $\diag(\uv)$ is a diagonal matrix with the diagonal elements in the vector $\uv$. Finally, we use the Matlab-like notation $[\Um]_{:,j}$ and $[\Um]_{i, :}$ to indicate the $j$-th column and the $i$-th row of a matrix $\Um$, and $[n_1:n_2]$ denotes the row vector of the integers from $n_1$ to $n_2$.

\section{System Model}
\label{sec:system_model}

This section summarizes the main concepts of the AMAF-RIS architecture from our previous work \cite{NewOldIdea} and extends the channel model to incorporate multipath propagation. The goal is to establish notation and context for the new contributions in this work, with detailed derivations available in prior publications.

\subsection{AMAF-RIS Module and Pragmatic PEM Design}
\label{subsec:amaf_ris_recap}

\begin{figure}[h!] 
\centerline{\includegraphics[width=8cm]{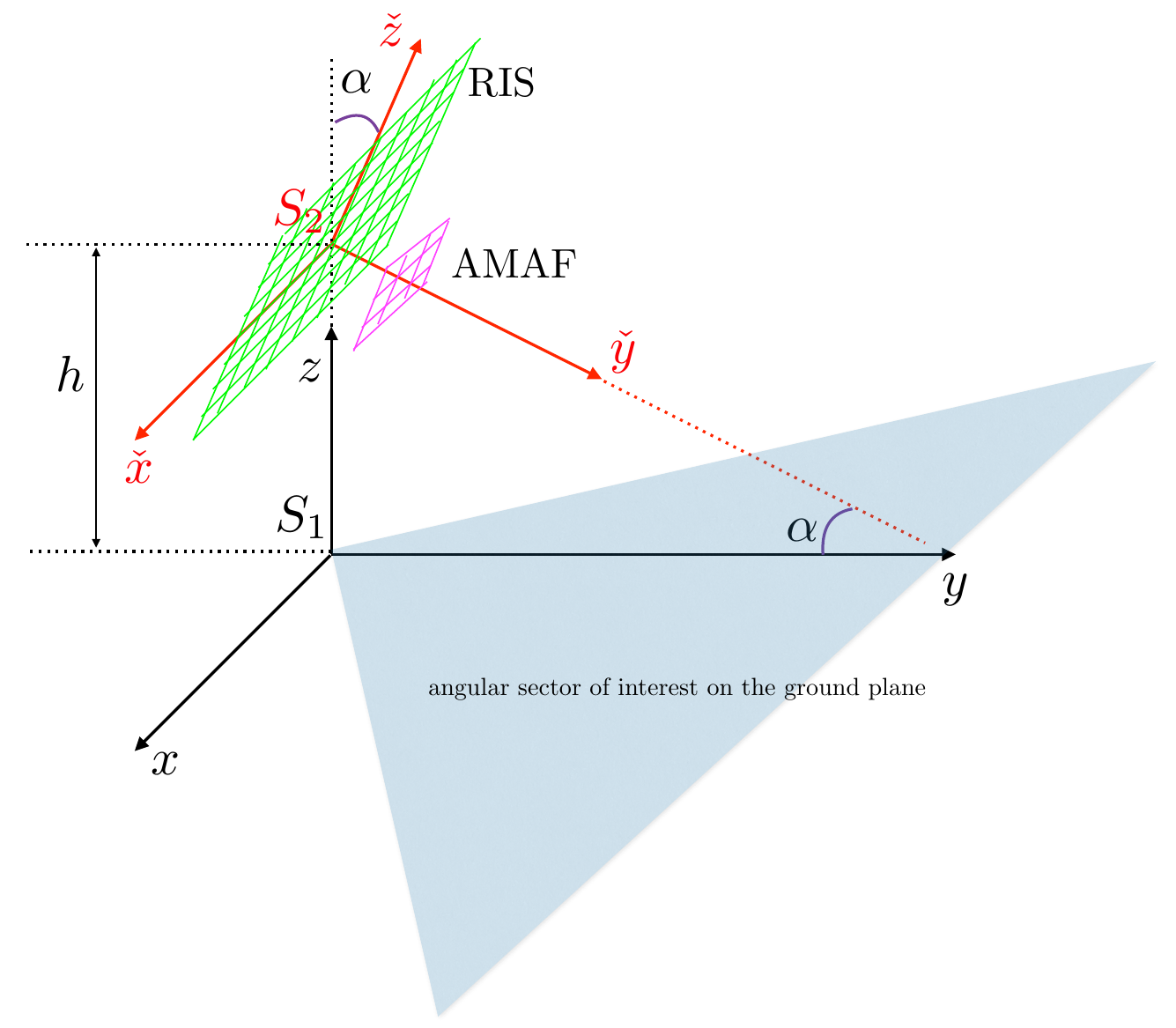}}
\caption{Array-fed RIS \cite[Fig. 1]{NewOldIdea}.}
\label{3D-fig11}
\end{figure}

\begin{figure}[h!] 
\centerline{\includegraphics[width=8cm]{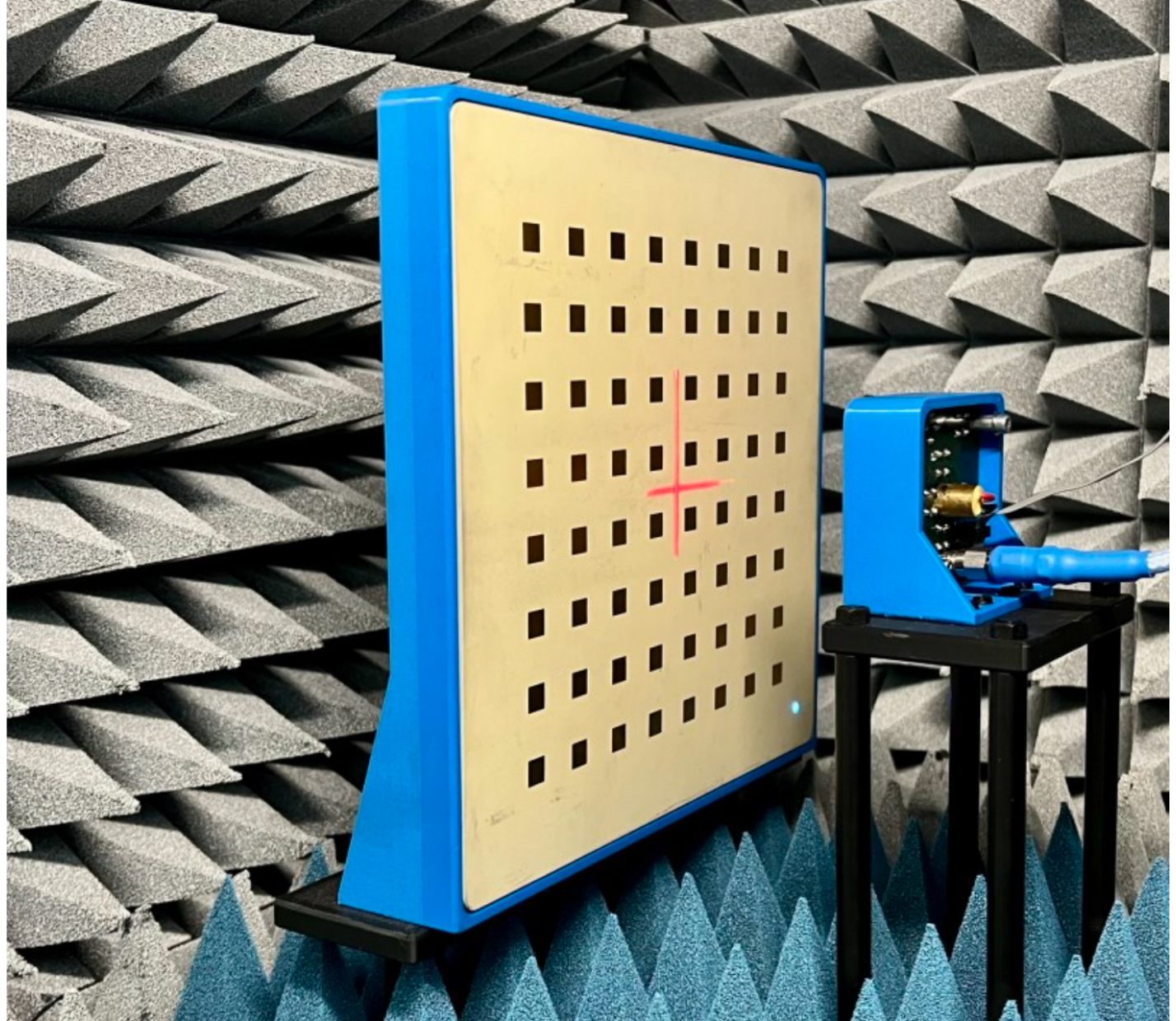}}
\caption{Array-fed RIS hardware \cite[Fig. 1]{hardware}.}
\label{fig:hardware}
\end{figure}

The core component of our architecture is a basic module
as shown in Fig. \ref{3D-fig11}, comprising a small active multi-antenna feeder (AMAF) placed in the near-field of a much larger passive reconfigurable intelligent surface (RIS). The AMAF is a square planar array of $N_a^2$ active elements (e.g., $N_a = 2$ for a $2 \times 2$ array), and the RIS is a square planar array of $N_p^2$ passive, controllable reflecting elements.\footnote{The generalization to rectangular non-square arrays is completely trivial and it is left to the reader.}
We consider a given carrier frequency $f_0$ with corresponding wavelength $\lambda_0 = c/f_0$ and signal bandwidth $W$ significantly smaller than $f_0$ (see \cite{NewOldIdea}). 
The arrays elements of AMAF and RIS are identical patches separated by $\lambda_0/2$ in both the horizontal and vertical direction, as described in \cite{NewOldIdea}, and all distances are given as multiples of $\lambda_0/2$. The patch radiation pattern is axisymmetric and frequency independent over the signal bandwidth, and it is given by $E(\phi, \theta) = 4 (\cos(\phi)\cos(\theta))^2$ for azimuth and elevation angles $\phi$ and $\theta$ with respect to the patch boresight direction. 

The antenna coefficients of the AMAF are fixed and depend only on the AMAF-RIS geometry and radiating element features. Let $f \in [-W/2, W/2]$ denotes the (baseband) frequency, and  let $\Tm(f) \in \mathbb{C}^{N^2_p \times N^2_a}$ be the frequency-dependent (baseband) near-field propagation matrix between the AMAF and RIS, whose entries are given by the Friis transmission formula including the patch radiation patterns \cite{NewOldIdea}.~\footnote{This was validated by accurate full-wave simulations in \cite{FWValidation}, \cite{APS} and also hardware in \cite{hardware}, see Fig. \ref{fig:hardware}.} We fix the AMAF coefficient vector $\bv \in \CC^{N_a\times 1}$ to be the principal eigenmode (PEM) of the AMAF-RIS transmission matrix at the center of the bandwidth,  i.e., we let $\bv = \vv_1$, where $\vv_1$ is the first right-singular vector of $\Tm(0)$. This maximizes the power transfer from the AMAF to the RIS for a given geometry at the center bandwidth. The resulting complex amplitude profile induced on the RIS is $\uv_1(f) = \Tm(f)\vv_1$, and this coincides at $f = 0$ with the principal left eigenmode of $\Tm(0)$.

The beamforming functionality is achieved solely by controlling the RIS elements phase shifts.
Let $\Um_1(f) = \vec^{-1}(\uv_1(f))$ denote the complex amplitude profile of the RIS elements at 
frequency $f$, induced by the PEM coefficients at the AMAF, formatted as the $N_p \times N_p$ array. 
We also define the array of phasors $\widetilde{\Wm} = \vec^{-1}(e^{-j\angle \uv_1(0)})$, the array of RIS beam shaping phase control $\Wm$, and the separable RIS beam steering to an azimuth/elevation angle of departure (AoD) $(\phi_0, \theta_0)$
\begin{equation} 
\Am(\phi_0,\theta_0; f)  = \av(\phi_0;f) \times \av(\theta_0;f), \label{steering}
\end{equation}
where we define
\begin{equation}
\av(\psi;f) := \exp\left ( -j\pi \frac{\lambda_0}{\lambda} [0:N_p-1] \sin \psi \right )^\transp,    \label{ULAsteering}
\end{equation}
as the steering vector at frequency $f$ for a standard uniform linear array (ULA) with $\lambda_0/2$ spaced elements, where $\lambda = c/(f_0 + f)$ for $f \in [-W/2,W/2]$.
Overall, the array of coefficients at the RIS elements at frequency $f$ is given by $\Xim \odot \Um_1(f)$, where
\begin{equation}  \label{eq:analogBFdesign}
\Xim = \Am(\phi_0,\theta_0;0) \odot \Wm \odot \widetilde{\Wm},
\end{equation}
is the array of RIS phase control coefficients. Notice that $\Um_1(f)$ and $\widetilde{\Wm}$ are entirely determined by the geometry of the AMAF-RIS module and by the AMAF PEM configuration, which are fixed. On the other hand, the array of phasors 
$\Am(\phi_0,\theta_0;0) \odot \Wm$ yields the {\em dynamic} phase control, where $\Wm$ is designed to create a desired {\em template beam shape} in the RIS boresight direction, and $\Am(\phi_0,\theta_0;0)$ determines the beam steering of the template beam around the AoD $(\phi_0, \theta_0)$.
The design of $\Wm$ is a main contribution of this work and it treated in Section~\ref{sec:flat_top_codebooks}. 

Notice also that the role of $\widetilde{\Wm}$ is to make the elements of $\widetilde{\Wm} \odot \Um_1(f)$ all real and non-negative 
$f = 0$. For later use, we define the {\em amplitude profile} at the center bandwidth induced on the RIS elements by the PEM configuration as 
$|\Um_1| := \widetilde{\Wm} \odot \Um_1(0)$. 

This design is ``pragmatic'' because the AMAF weights are hard-wired, determined only by the fixed geometry, and set to maximize the power transfer between the AMAF and the RIS at the center bandwidth. However, in this way the need for costly vector modulators at the AMAF is eliminated. 
The resulting PEM beam is highly directional with very low sidelobes due to the natural aperture taper of $|\Um_1|$ as shown in \cite[Fig. 7]{NewOldIdea}. 
Furthermore, notice that the RIS phase control $\Xim$ in \eqref{eq:analogBFdesign} is independent of frequency, and this necessarily incurs some beam squint effect. However, 
we have already shown in \cite{NewOldIdea} that these mismatch effects 
in terms of beam shape distortion for $f \in [-W/2, W/2]$ are essentially negligible for $W \leq 0.05 \cdot f_0$.

\subsection{Multi-Module MU-MIMO Architecture}
\label{subsec:multi_module}

To serve up to $K$ users via spatial multiplexing \cite{JSDM}, we employ the one-stream-per-subarray (OSPS) paradigm \cite{commit_bf1}. This is realized by stacking $K$ identical AMAF-RIS modules. Each module is driven by a single RF chain and is responsible for forming one data stream towards its intended user.

The $K$-module system introduces two potential interference sources:
Near-End Cross-Talk (NEXT): Signal leakage from one AMAF to the RIS of another, adjacent module.
Far-End Cross-Talk (FEXT): Interference at a user caused by the sidelobes (or, in this work, multipath reflections) of beams intended for other users.
Let $\Tm_{\ell,j}(f) \in \CC^{N^2_p \times N^2_a}$ denote the near-field propagation matrix 
between the AMAF of the $j$-th module and the RIS of the $\ell$-module, where 
$\Tm_{\ell,\ell}(f)$ coincides with $\Tm(f)$ in Section \ref{subsec:amaf_ris_recap} because the modules are identical, and let $\Xim_\ell$ denote the RIS phase control of the $\ell$-th module as defined in \eqref{eq:analogBFdesign}. For a given set of $K$ users, let $\gv_{k,\ell}(f) \in \CC^{N^2_p \times 1}$ denote the
propagation channel from the RIS of the $\ell$-th module and user $k$ (assumed with a single antenna).
Because of multipath, $\gv_{k,\ell}(f)$ is generally frequency-selective and therefore depends on $f$.
Then, the effective channel at frequency $f$ between the $K$ AMAF antenna ports and the $K$ users 
is a matrix $\Hm(f) \in \CC^{K \times K}$ with $(k,j)$ element given by 
\begin{equation}
[\Hm(f)]_{k,j} = \sum_{\ell=1}^K \gv_{k,\ell}(f)^\herm \diag(\vec(\Xim_\ell) \Tm_{\ell,j}(f) \vv_1. \label{effective channel}
\end{equation}
In \cite{NewOldIdea}, letting $F/D$ denote the ratio of the AMAF-RIS distance to the RIS linear size,\footnote{This will be referred to as normalized focal length in the following.} 
we showed that for $F/D$ not larger than 0.6 the NEXT is fully negligible even when the modules are separated by $\lambda_0/2$. Under this condition,
the blocks $\Tm_{\ell,j}(f)$ for $\ell \neq j$ are essentially zero, such that $\Hm_{k,j}(f)$ in \eqref{effective channel} reduces to 
\begin{equation}
[\Hm(f)]_{k,j} \approx \gv_{k,j}(f)^\herm \diag(\vec(\Xim_j) \Tm_{j,j}(f) \vv_1. \label{effective channel-noNEXT}
\end{equation}
Notice that NEXT depends uniquely by the geometry of the basic module and on the staking of the $K$ modules, and not on the propagation channels. 

We also showed that for pure LOS propagation (i.e., when the channels $\gv_{k,\ell}(f)$ are just array steering vectors), by choosing groups of $K$ users with sufficient angular separation, also the FEXT is essentially negligible with respect to the useful signal term, i.e., with an appropriate RIS phase control,  the off-diagonal terms $|\gv_{k,j}(f)^\herm \diag(\vec(\Xim_j) \Tm_{j,j}(f) \vv_1|^2$ are much smaller than diagonal terms $|\gv_{k,k}(f)^\herm \diag(\vec(\Xim_k) \Tm_{k,k}(f) \vv_1|^2$. 
As we will show, this is not the case in multipath environments.

\subsection{Multipath Channel Model with vMF Scatterers}
\label{subsec:multipath_model}

In this paper we consider a more realistic multipath environment where the physical channel vector $\gv_{k,\ell}(f)$ is given by 
\begin{equation}
\gv_{k,\ell}(f) = \gv_{k,\ell}^{\text{LOS}}(f) + 
\underbrace{\sum_{s=1}^{S} \gv_{k,\ell}^{(s)}(f)}_{\mbox{NLOS component}}.
\label{eq:total_channel}
\end{equation}
Consider Fig.~\ref{3D-fig11}, where the RIS coordinate systems $S_2$ is a translated and down-tilted by an angle $\alpha$ version of the coordinate system of the ground plane $S_1$. The channel LOS component is given by 
\begin{eqnarray}
\gv_{k,\ell}^{\text{LOS}}(f) & = & {\sqrt{L_k^{\text{LOS}} \; E(\phi_k^\text{LOS},\theta_k^\text{LOS})}} \, e^{-j2\pi f \tau_k^\text{LOS}} \nonumber \\
& &  e^{-j \pi \frac{\lambda_0}{\lambda} \widehat{\nv} \cdot \pv_\ell} \vec \left ( \Am( \phi_k^\text{LOS}, \theta_k^\text{LOS};f) \right ) 
\label{eq:los_channel}
\end{eqnarray}
where $L_k^{\text{LOS}}$ is pathloss coefficient, $\phi_k^\text{LOS}, \theta_k^\text{LOS}$ is the LOS path AoD, 
$\tau_k^\text{LOS}$ is the delay from the BS to user $k$, 
$\widehat{\nv}$ is the normal unit vector of the planar wave departing from the BS plane with AoD $\phi_k^\text{LOS}, \theta_k^\text{LOS}$ and 
$\pv_\ell$ is the coordinate of the center $\ell$-th basic module with respect to the BS coordinate system, normalized by $\lambda_0/2$.\footnote{The path loss exponent for both the LOS and the specular NLOS components is set to 2, corresponding to free-space propagation. This is a common modeling assumption for dominant specular reflections, as they do not experience the same scattering and diffraction losses as diffuse multipath components, which typically require larger exponents.}
Notice that the steering vector $\Am(\phi,\theta; f)$ and the relative $\ell$-th module phase displacement term $e^{-j \pi \frac{\lambda_0}{\lambda} \widehat{\nv} \cdot \pv_\ell}$ in \eqref{steering}
are frequency-dependent. Hence, in our results, the effects of beam-squint are fully taken into account. 

For the NLOS components modeling, each scatterer $s$ is characterized by
an AoD $(\phi^{(s)}, \theta^{(s)})$ with respect to the BS coordinate 
system $S_2$, that depends only on the relative position of the scatterer with respect to the BS, a path delay $\tau^{(s)}_k$ with respect to user $k$, and a complex path amplitude $\gamma^{(s)}_k$, incorporating path loss, the radar cross-section of the $s$-th scatterer and a 
uniformly distributed phase due to the fact that the path phase rotation
$2\pi [ \tau^{(s)}_k f_0 ]_{{\rm mod} [0,1]}$, when 
$\tau^{(s)}_k f_0$ is a large number 
equal to the propagation distance divided by the wavelength, is essentially uniformly distributed on $[0,2\pi]$.

To model realistic, non-isotropic scattering, we generate the scatterers' spatial directions using the von Mises-Fisher (vMF) distribution in 3D \cite{MarzettaVMF, JohnVMF}. The vMF distribution, a spherical analogue of the Gaussian, is defined by a mean direction $\mv_c$ and a concentration parameter $\kappa_c$. The probability density function for a unit vector $\rv$ is:
\begin{equation}
p(\rv \mid \mv_c, \kappa_c) = C_p(\kappa_c) \exp(\kappa_c \, \mv_c^{\transp} \rv), \forall c = 1,2,\twodots,C,
\label{eq:vmd_distribution}
\end{equation}
where $C$ is the total number of clusters, $C_p(\kappa)$ is the normalizing constant on the unit sphere in $\mathbb{R}^3$.  
We impose the physical constraint that only scatterers in the front hemisphere of the RIS are accepted:
\begin{equation}
\rv^{(s)} \sim \text{vMF}(\mv_c,\kappa_c)\,\mathbbm{1}\!\left\{ \rv^{(s)} \cdot \check{\jv} > 0 \right\},
\end{equation}
where $\check{\jv}$ is the RIS boresight vector. 
This is motivated by the fact that the radiation patter $E(\phi,\theta)$ 
of the RIS elements approaches zero at the hemisphere boundary, and the RIS reflects (and therefore transmits) only in the front hemisphere.

A large $\kappa_c$ implies scatterers are tightly clustered around $\mv_c$, modeling the case of a dominant reflecting object.  
For the system described in Section~\ref{sec:system}, with the base station height $h = 20$m
and the mechanical downtilt angle $\alpha = 37.37 \degree$, see Fig.~\ref{3D-fig11} and \cite[Section V]{NewOldIdea}, the scatterer clusters lie within elevation angles from $-26^\circ$ to $+26^\circ$ and azimuth angles from $-60^\circ$ to $+60^\circ$, which places all mean directions well inside the RIS front hemisphere. 

After generating the path directions as explained above, we assign a range to each accepted vMF path such that the scatterers occupy realistic 3D positions within a specific height range above the ground plane in the coordinate system $S_1$. The delay $\tau^{(s)}_k$ is determined 
by the length of the one-reflection path from the BS to the scattering element $s$ and then to user $k$ on the ground.

The NLOS channel from module $\ell$ to user $k$ via scatterer $s$ is given by
\begin{eqnarray}
\gv^{(s)}_{k,\ell}(f) & = & {\sqrt{E(\phi^{(s)},\theta^{(s)})}} \gamma^{(s)}_k \, e^{-j2\pi f \tau^{(s)}_k} \nonumber \\
& & e^{-j \pi \frac{\lambda_0}{\lambda} \widehat{\nv} \cdot \pv_\ell} \vec \left ( \Am( \phi^{(s)}, \theta^{(s)};f) \right ),
\label{eq:nlos_channel}
\end{eqnarray}
where $\widehat{\nv}$ has the same meaning as defined before, for 
AoD $(\phi^{(s)},\theta^{(s)})$.


Notice that the $S$ scatterers are common to all $K$ users and therefore create not nearly mutually orthogonal channel vectors, despite the LOS components may have sufficient angular separation such that they are nearly mutually orthogonal. This geometrically consistent channel model allows the accurate evaluate the system performance in the presence of structured multipath, which significantly increases the FEXT between users (as we'll see in Section \ref{sec:results}).

\subsection{Users-to-beam assignment}
\label{subsec:rsrp_computation}

We consider a fixed-size and pre-designed beamforming codebook defined by the set of RIS phase control arrays $\{\Xim^{(1)}, \ldots, \Xim^{(C)}\}$.
This codebook is common to all $K$ basic modules forming the BS. 
A population of $U$ users is placed in the coverage area on the ground plane 
(e.g., a sector as in Fig.~\ref{3D-fig11}). The BS performs periodically a beam acquisition scheme where mutually orthogonal  reference signals 
are sent through the codebook beams, and users can identify their ``best beam'' from the received signal.  
Since the $K$ modules are identical and essentially co-located at the BS, 
each module can probe in parallel up to $C/K$ beamforming codewords, 
using repeatedly one of $K$ mutually orthogonal reference signals.
However, other schemes are possible that use less Tx power by activating 
less than $K$ modules for a longer time. Here we do not discuss these 
details and assume that, at the end of the beam acquisition phase, each 
user has measured the received signal power associated to all $C$ beamforming codewords. Then, the users notify the BS of their selected beam index via a control channel in the uplink.  The user-beam association is performed periodically and at relatively low duty cycle, because the user position, especially in a mmWave system with nomadic users, changes relatively slowly.\footnote{Think of a local mmWave hotspot designed to achieve high throughput in low mobility, e.g., serving a high density area such as an airport hall.} In this work, we choose as criterion for beam selection the maximum reference signal received power (RSRP). 
For user $k$ and beam $c$, the {\em wideband} RSRP is given by
\begin{equation}
G_k^{(c)} = \sum_{\nu=1}^{N_{\text{sub}}} 
\left| \gv_{k,1}(f_\nu)^\herm \, \diag{(\vec(\Xim^{(c)}))} \, \Tm_{1,1}(f_\nu) \, \vv_1 \right|^2,
\label{eq:rsrp}
\end{equation}
where $N_{\text{sub}}$ is the number of OFDM subcarriers, $f_\nu$ is the $\nu$-th discrete subcarrier,  and without loss of generality we have assumed that the codebook is transmitted from module 1, since the modules are identical and transmit mutually orthogonal reference signals (e.g., a set of mutually orthogonal Zadoff-Chu (ZC) sequences \cite{Book5GNR,SRSref, TR38802}). Each user $k$ is then assigned to the beam achieving the maximum RSRP, i.e., to the beam with index $c_k = \argmax\{G^{(c)}_k : c \in [1:C]\} $. Notice that the RSRP can be accurately measured provided that the reference signals span sufficiently high dimensions, such that noise averaging is effective. In addition, the averaging over the OFDM subcarriers in \eqref{eq:rsrp} ensures that user association reflects wideband performance rather than a single-subcarrier snapshot, providing robustness against frequency-selective fading in multipath environments.

\section{Hierarchical Codebook with Flat-Top Beams}
\label{sec:flat_top_codebooks}

Having established the system model, in this section we consider the design of the beamforming codebook. 
As anticipated in Section \ref{subsec:amaf_ris_recap}, the RIS phase control 
of the basic module $\Xim$ in \eqref{eq:analogBFdesign} is formed by 
the elementwise product (or sum in the phase domain) of $\widetilde{\Wm}$, which is fixed by the PEM configuration of the AMAF and by the near-field AMAF-RIS propagation matrix $\Tm(0)$, which in turn depends only on the 
geometric shape of AMAF, RIS, and $F/D$ ratio, the steering linear phase gradient $\Am(\phi_0,\theta_0;0)$ in the direction of a desired AoD, and the 
beam shaping unit modulus coefficients $\Wm$, which determine the shape of the beams. Our design strategy consists of optimizing a desired beam shape
in terms of flat-top beam width and side lobe rejection, for the boresight direction angle $\phi_0 = \theta_0 = 0$. Eventually, the codebook $\{\Xim^{(c)} : c \in [1:C]\}$ is obtained by combining suitably designed beam shapes (defined by $\Wm$) and center beam steering
angles determining $\Am(\phi_c,\theta_c;0)$ for angles
$(\phi_c, \theta_c) : c \in [1:C]$, 
in order to provide a good coverage of the desired area. 

Since the RIS phase-shifters are frequency-invariant, we design the shape of the template beam at $f = 0$ (center bandwidth) and keep these coefficients for the whole signal bandwidth, thus incurring in some beam-shape distortion at the bandwidth edges \cite{NewOldIdea}. As said before, the numerical results presented in this paper take into  full account the frequency dependency of the AMAF-RIS matrix $\Tm(f)$ and the propagation channels $\gv_{k,\ell}(f)$, i.e., no approximation is made in the simulations.

\subsection{Pragmatic Flat-Top Beam Design}
\label{subsec:flat_top_design}

\begin{figure*}[ht] \vspace{-0.3cm}
    \centering
    \begin{subfigure}[b]{0.495\textwidth}
        \includegraphics[width=8cm,height=6cm]{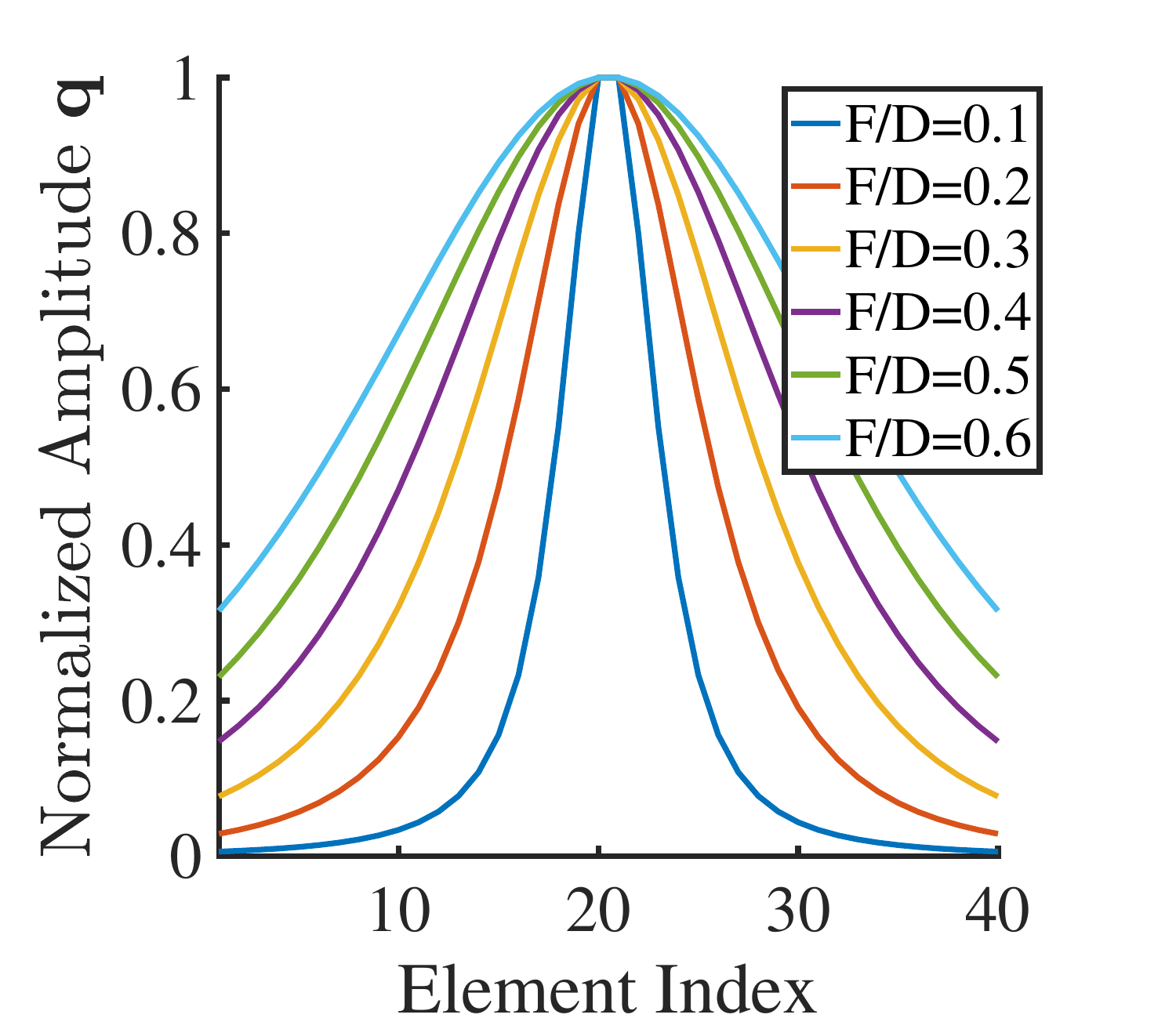}
        \caption{Antenna domain}
        \label{fig:marginalsAntenna}
    \end{subfigure}
    \hfill
    \begin{subfigure}[b]{0.495\textwidth}
        \includegraphics[width=8cm,height=6cm]{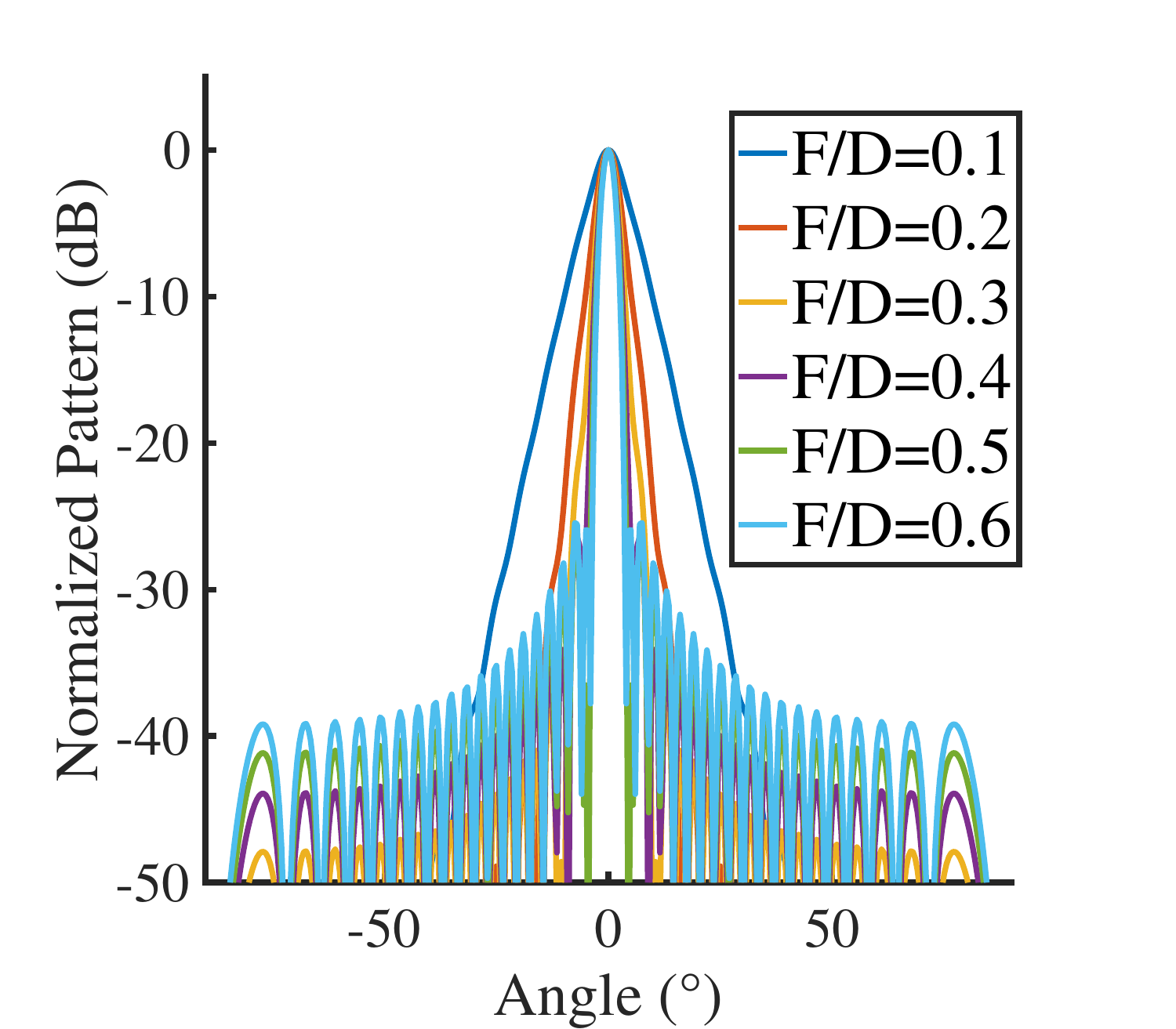}
        \caption{Far-field radiation patterns}
        \label{fig:marginalsFarField}
    \end{subfigure}

    \caption{Marginal distributions of 2D $|\Um_1|$.}
    \label{fig:marginals}
\end{figure*}

Recall that the {\em amplitude profile} at the center bandwidth induced on the RIS elements by the PEM configuration as $|\Um_1| := \widetilde{\Wm} \odot \Um_1(0)$.
Motivated by the goal of reducing the dimensionality, we first approximate $|\Um_1|$ by a separable function, i.e., we seek the best approximation 
$|\Um_1| \approx \qv \times \qv$. Then, we develop a beam pattern design approach for the 1D uniform linear array (ULA) case with amplitude profile $\qv$. Finally, we obtain the phase control coefficients of the 2D case as the Cartesian product of the 1D phase coefficients.  This procedure is in agreement with the well-known approach in digital filter theory 
to design 2D filters as separable compositions of 1D filters. 

\subsubsection{Best separable approximation}

Without loss of generality, we can consider $|\Um_1|$ to be normalized to have element sum equal to 1. An $N_p \times N_p$ array with non-negative elements that sum to 1 defines a joint probability mass function (pmf) on the 
integer-valued square $[1:N_p] \times [1:N_p]$.  
In order to find the best separable approximation and quantify the incurred approximation error, we seek the 1D pmf $\qv$ that minimizes the Kullback-Leibler (KL) divergence 
\begin{equation}
D(|\Um_1| \; \| \; \qv \times \qv) := \sum_{i=1}^{N_p} \sum_{j=1}^{N_p} \log \frac{[|\Um_1|]_{i,j}}{q_i q_j}. \label{KL}
\end{equation}
It is well-known (see \cite{Cover:2006}) that $D(|\Um_1| \| \qv \times \qv)$
is minimized by letting
\begin{eqnarray}
\qv & = & \sum_{j=1}^{N_p} [|\Um_1|]_{:,j} = \sum_{i=1}^{N_p} [|\Um_1|]_{i,:}^\transp,  \label{optq}
\end{eqnarray}
i.e., the optimal $\qv$ is the marginal distribution of $|\Um_1|$.\footnote{
When both the AMAF and the RIS are square arrays and they are co-axially placed, the symmetry of the propagation yields that $|\Um_1|$ is 
a symmetric matrix and therefore the row-sum and the column-sum in \eqref{optq} coincide. The approach can be easily generalized to the non-symmetric case, where in this case the optimal separable approximation is in the form $\qv_x \times \qv_z$ where $\qv_x$ and $\qv_z$ are the $x$-marginal (column sum) and $z$-marginal (row sum) pmfs of $|\Um_1|$.}
To quantify the level of separability, we consider the information theoretic identity
\[ D(|\Um_1| \; \| \; \qv \times \qv) = 2 H(\qv) - H(|\Um_1|) \]
where $H(P)$ indicates the entropy of the pmf $P$ \cite{Cover:2006}. 
From the elementary properties of entropy and KL divergence \cite{Cover:2006}, we have that 
$D(|\Um_1| \; \| \; \qv \times \qv) \leq 2 H(\qv)$ and that 
$D(|\Um_1| \; \| \; \qv \times \qv) = 0$ if and only if 
$|\Um_1| = \qv \times \qv$, i.e., when $|\Um_1|$ is {\em exactly} separable.
Hence, a meaningful quantity to define the level of separability is 
\[ \eta = \frac{D(|\Um_1| \; \| \; \qv \times \qv)}{2 H(\qv)}, \]
such that $0 \leq \eta \leq 1$ and $\eta = 0$ indicates exact separability. 

In this paper, we develop a specific design example for a basic module with AMAF of size $2 \times 2$ and RIS of size $40 \times 40$ (all elements are $\lambda_0/2$ spaced, see \cite{NewOldIdea}), briefly denoted in the following as AMAF$_2$-RIS$_{40}$. For fixed $N_p$ and $N_a$, the key parameter determining the amplitude profile $|\Um_1|$ is the normalized focal length $F/D$. For $F/D$ ranging from $0.1$ to $0.6$, we verified that the separability index $\eta$ is no larger
than $0.016$ (the largest value $0.016$ is at $F/D=0.1$, it decreases with increasing $F/D$, $\eta=0.007$ for the chosen $F/D=0.2$).
Fig.~\ref{fig:marginalsAntenna} shows the separable 1D amplitude profile $\qv$ for $F/D$ ranging from 0.1 to 0.6, and Fig.~\ref{fig:marginalsFarField} shows 
the corresponding ULA far-field radiation patterns in the angle domain.
We notice that as $F/D$ increases, the amplitude profile widens, and consequently the beam radiation pattern becomes more and more pointy. Since our goal is to create beams with a controllable width and flat-top main lobe, we shall choose a small value of $F/D$. This has also the advantage that the resulting AMAF-RIS basic module will have a compact form factor. 

\subsubsection{Beam widening by phase perturbation}

    \begin{figure}[h]
        \centering
        \includegraphics[width=6.75cm]{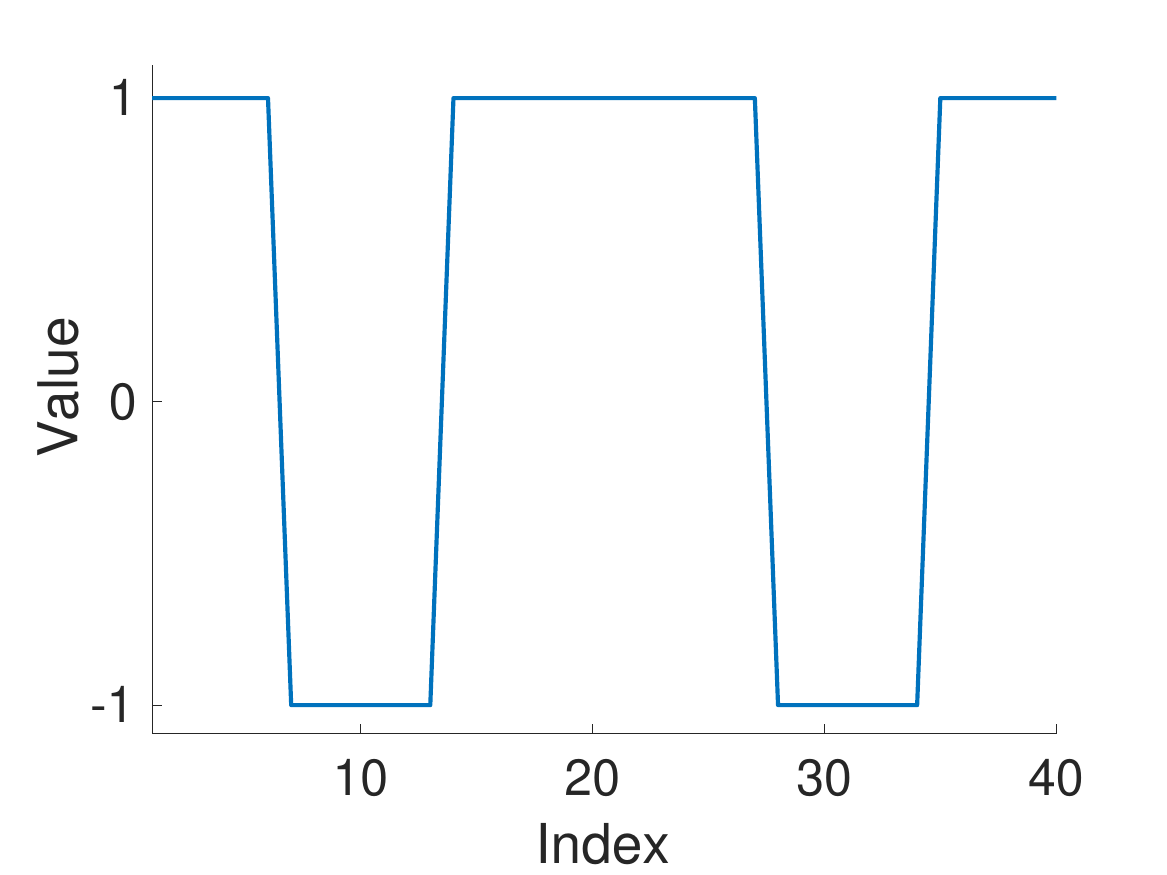}
        \caption{Binary phase profile.}
        \label{fig:BinaryPattern}        
    \end{figure}
    
\begin{figure*}[h!] \vspace{0cm}
    \centering
    \begin{subfigure}[b]{0.328\textwidth}
        \includegraphics[width=\linewidth]{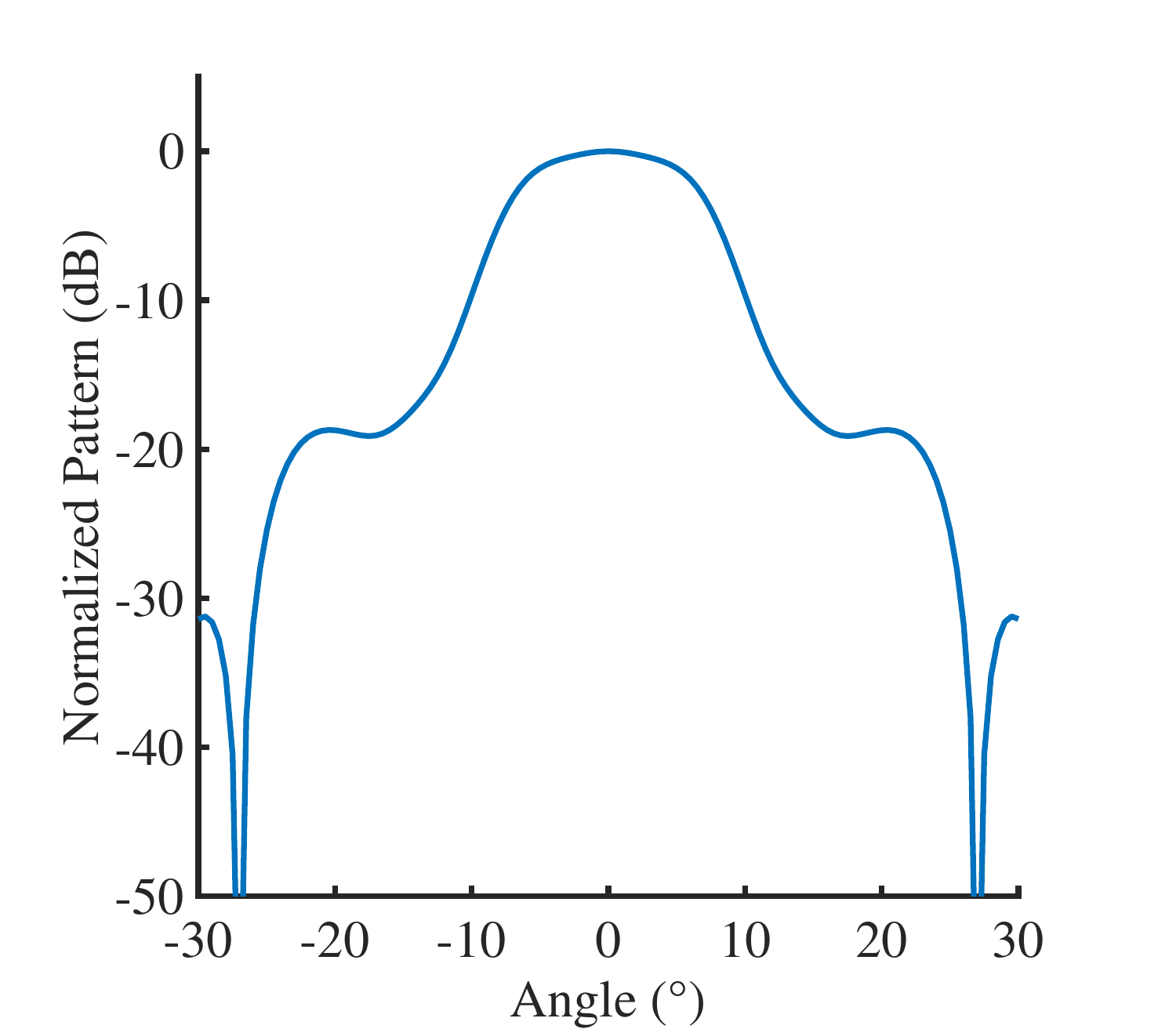}
        \caption{$F/D = 0.1$}
        \label{fig:FbyD1}
    \end{subfigure}
    \hfill
    \begin{subfigure}[b]{0.328\textwidth}
        \includegraphics[width=\linewidth]{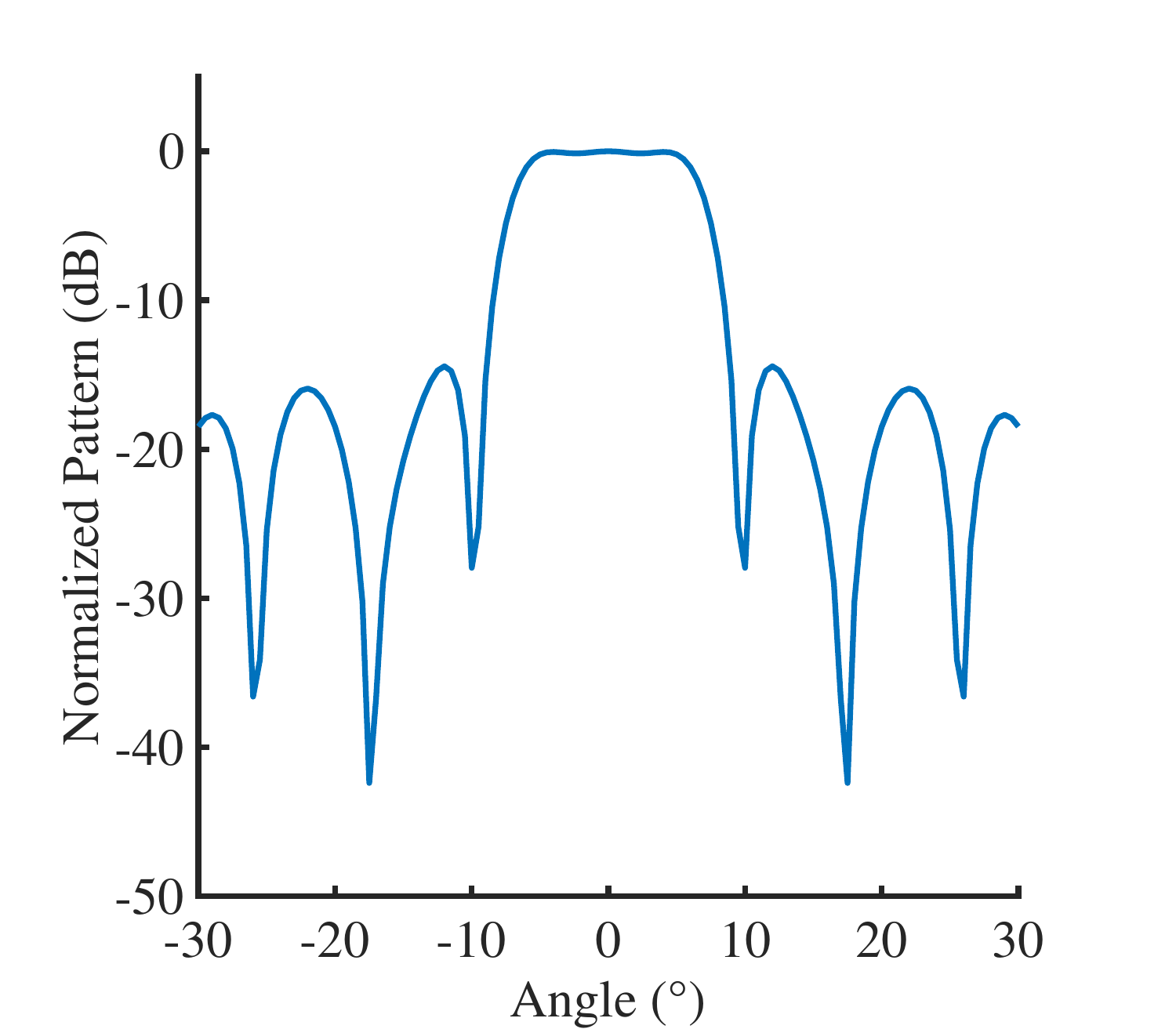}
        \caption{$F/D = 0.2$}
        \label{fig:FbyD2}
    \end{subfigure}
    \hfill
    \begin{subfigure}[b]{0.328\textwidth}
        \includegraphics[width=\linewidth]{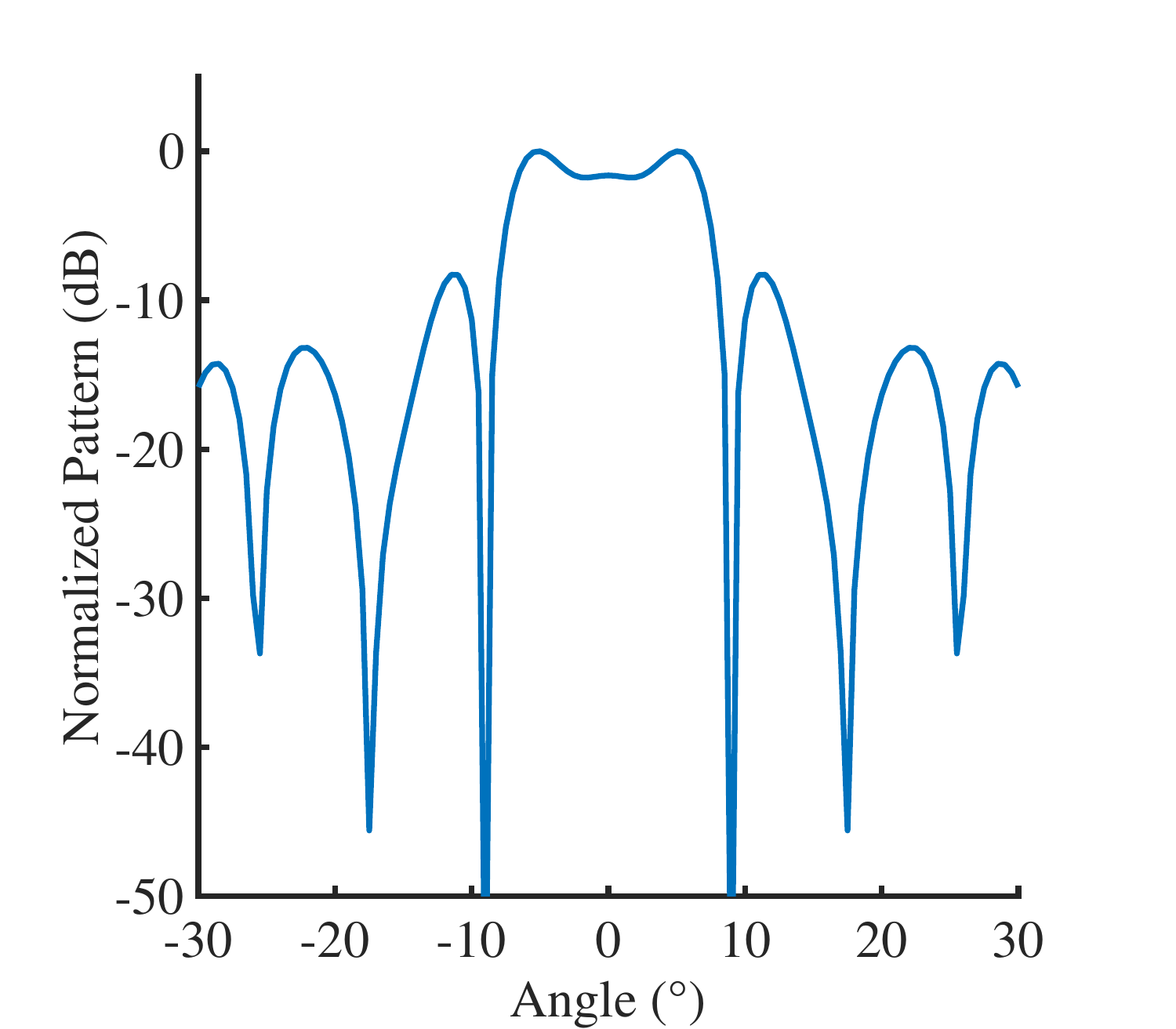}
        \caption{$F/D = 0.3$}
        \label{fig:FbyD3}
    \end{subfigure}
    
    \vspace{0.2cm}
    \begin{subfigure}[b]{0.328\textwidth}
        \includegraphics[width=\linewidth]{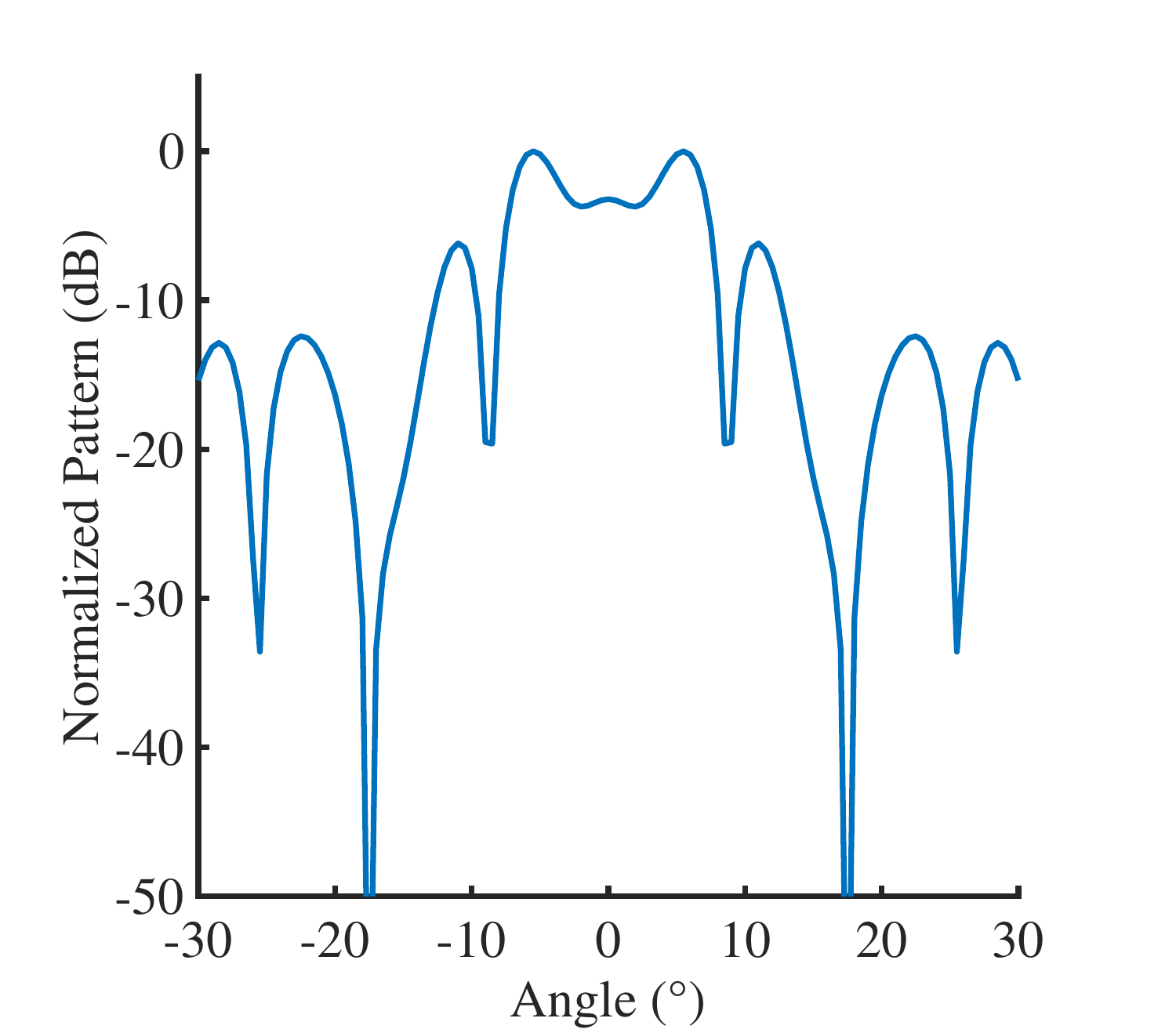}
        \caption{$F/D = 0.4$}
        \label{fig:FbyD4}
    \end{subfigure}
    \hfill
    \begin{subfigure}[b]{0.328\textwidth}
        \includegraphics[width=\linewidth]{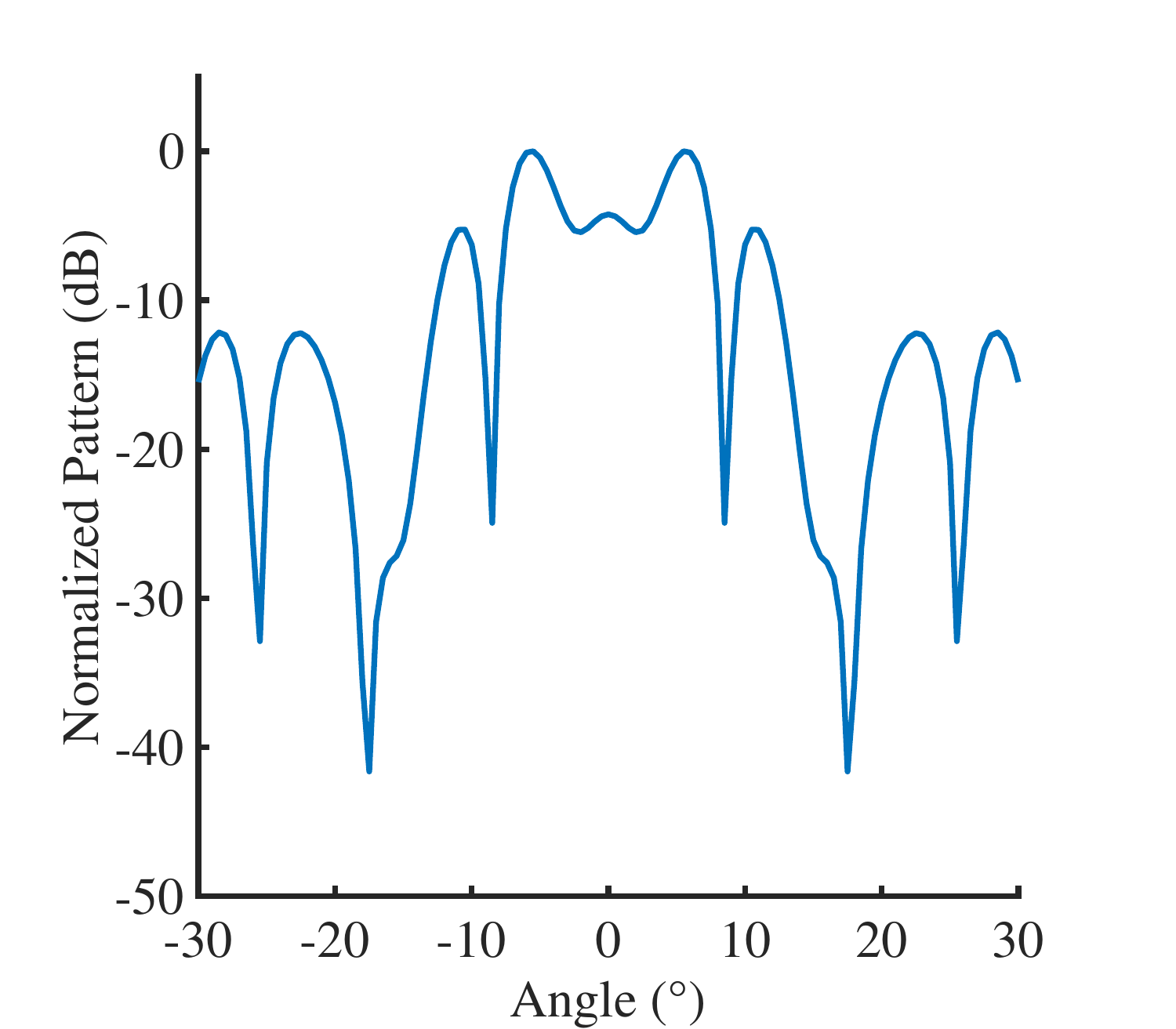}
        \caption{$F/D = 0.5$}
        \label{fig:FbyD5}
    \end{subfigure}
    \hfill
    \begin{subfigure}[b]{0.328\textwidth}
        \includegraphics[width=\linewidth]{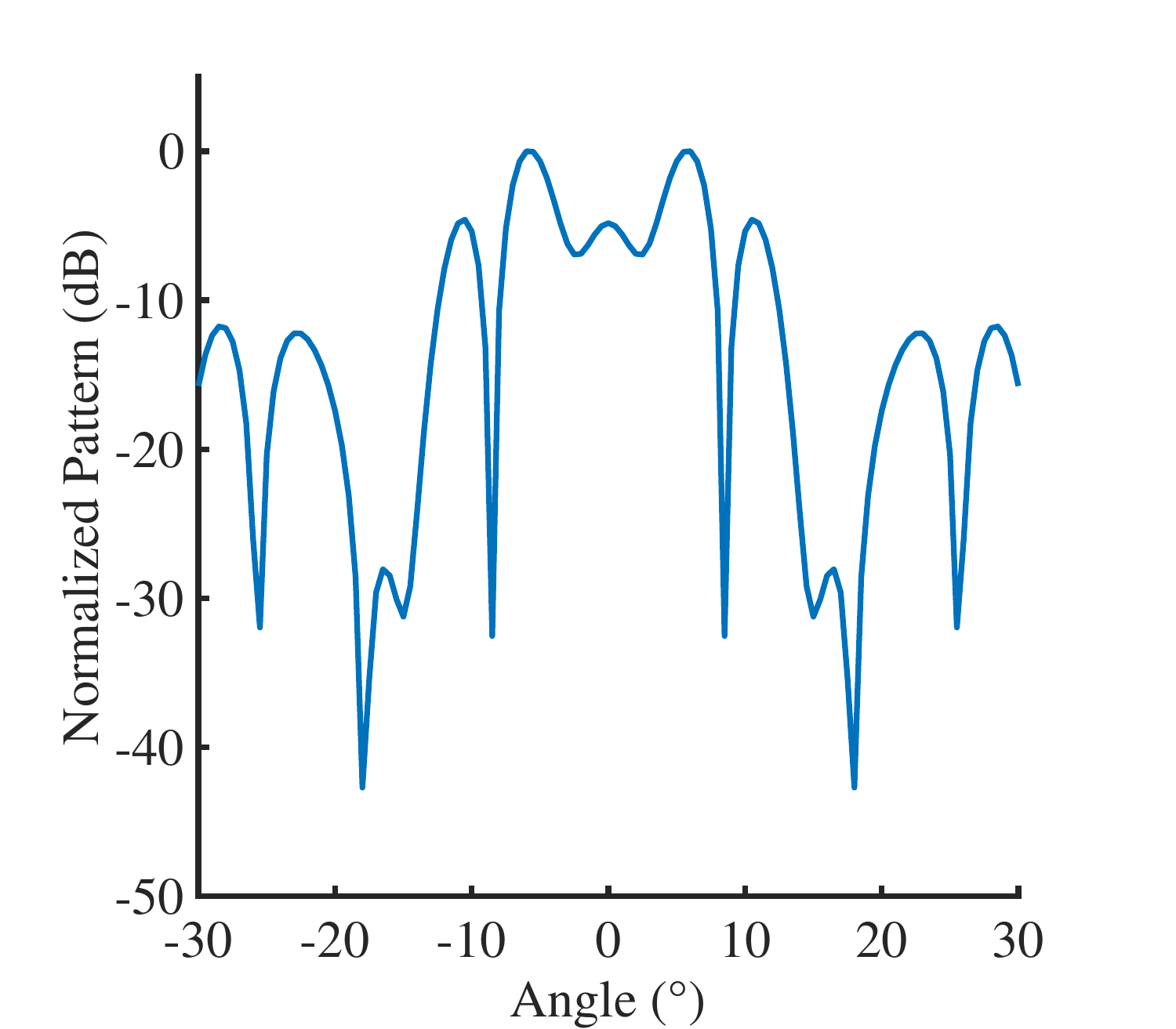}
        \caption{$F/D = 0.6$}
        \label{fig:FbyD6}
    \end{subfigure}
    
    \caption{$F/D$-wise far-field radiation patterns of an ULA with $N_p = 40$ antennas and amplitude profile $\qv$, multiplied by the binary phase profile of Fig.~\ref{fig:BinaryPattern}.}
    \label{fig:F_by_D_wise patterns}
\vspace{0cm} \end{figure*}

Owing to the formal analogy between the discrete Fourier transformation 
relating the time domain coefficients and the frequency response in digital FIR filters \cite{firpm,trees} and the relation between antenna coefficients and far-field radiation pattern in ULAs, we expect that, to achieve a smooth flat-top, steep transition, low-sidelobe beam shape, the antenna coefficients should mimic a ``sinc'' function, whose Fourier transform is a {\em brickwall function}. 
In particular, the antenna coefficient values, seen as a function of the discrete antenna index, should have a positive and symmetric center lobe, and blocks with alternating signs symmetrically arranged with respect to the center of the ULA. 
Motivated by this consideration, we adopt the 1D binary phase profile $\wv_{\rm binary}$ obtained from Parks-McClellan FIR filter design algorithm \cite{firpm}, as shown in Fig.~\ref{fig:BinaryPattern}. This pattern is applied elementwise to the ULA amplitude profile $\qv$ such that the real-valued (positive and negative) ULA coefficient vector is given by 
$\wv_{\rm binary} \odot \qv$.  The resulting radiation patterns for different values of $F/D$ are shown in  Fig.~\ref{fig:FbyD2}. In light of these results, for the design case at hand, we choose $F/D = 0.2$.

\begin{figure}[h!] 
    \centering
    \begin{subfigure}[b]{0.5\textwidth}
        \centering
        \includegraphics[width=\textwidth]{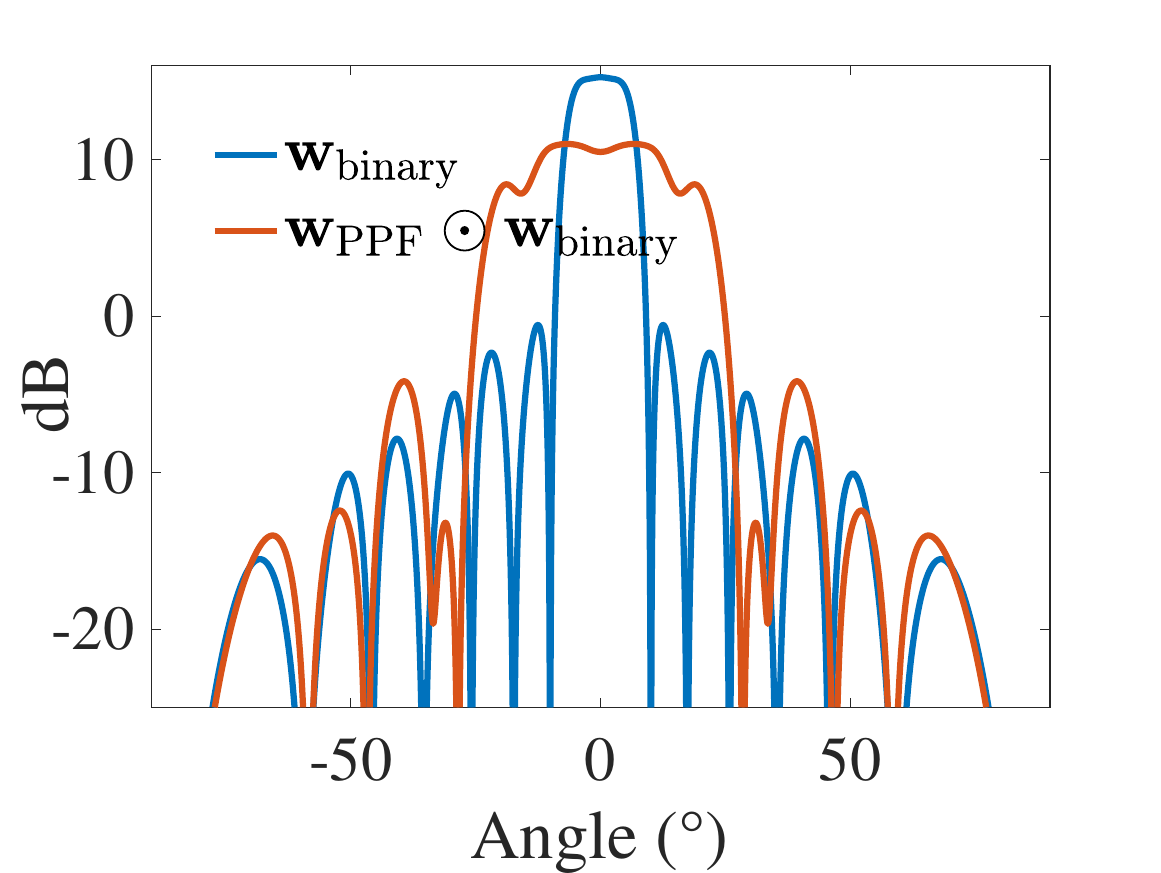}
        \caption{Flat-top beam patterns obtained from binary phases plus phase perturbation.}
        \label{fig:step2}
    \end{subfigure}
    
    \vspace{0.3cm} 

    \begin{subfigure}[b]{0.5\textwidth}
        \centering
        \includegraphics[width=\textwidth]{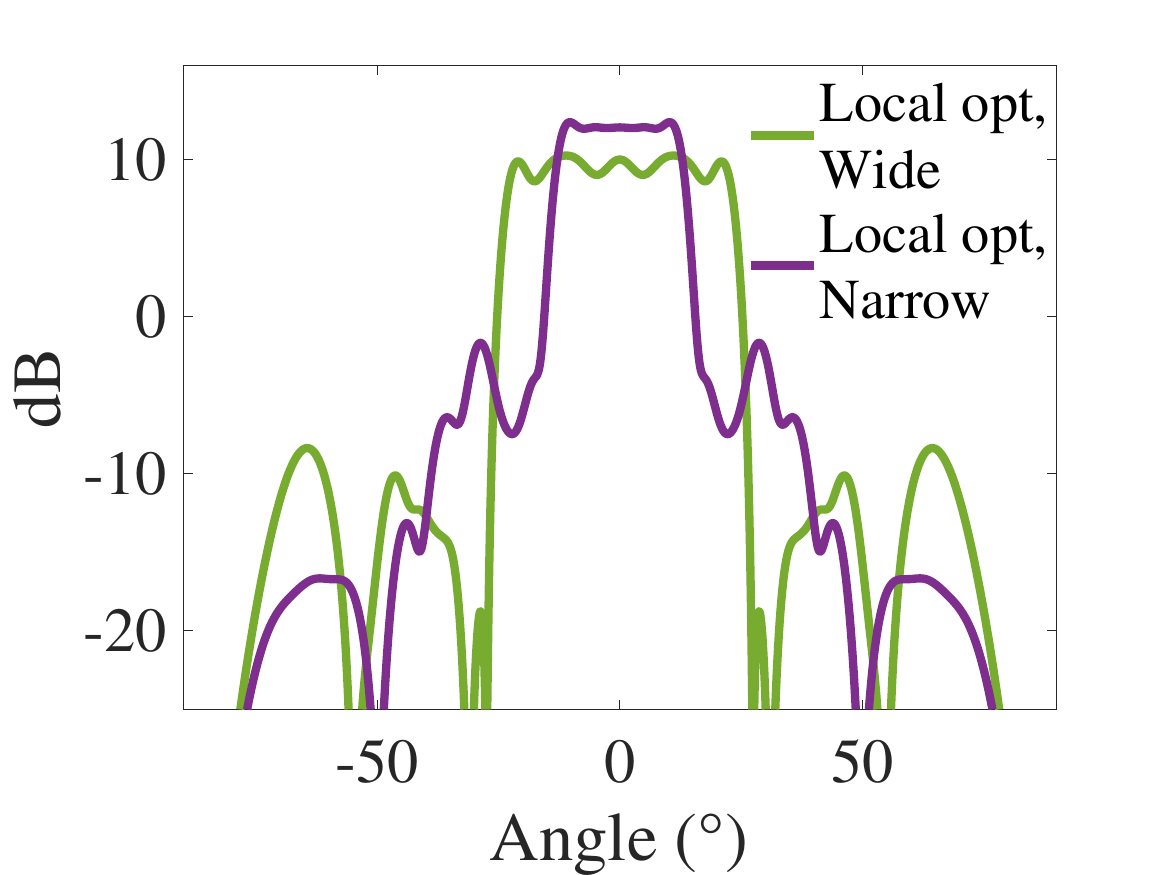}
        \caption{Tunable width flat-top beams obtained by local optimization, with the initialization $\wv_{\rm ini} = \wv_{\rm PPF} \odot \wv_{\rm binary}$.}
        \label{fig:step4}
    \end{subfigure}

    \caption{Comparison of beam patterns: (a) perturbed phases only; (b) local optimization with perturbed phases initialization.}
    \label{fig:combined}
\end{figure}  

In general, the width of the flat-top beam obtained from the binary phase described above is not widely tunable because the $F/D$ ratio is a mechanical form factor that cannot be adaptively changed. To achieve a tunable beamwidth, we apply a phase perturbation function (PPF) \cite{R1-1611929, R1-1700772} for beam widening:
    \begin{equation} 
        f(n) = \left| 4\pi \varrho \left( \frac{0.5}{N_p - 1} + \frac{n - 0.5N_p}{N_p - 1} \right)^\Pi \right|,
        \label{eq:ppf}
    \end{equation}
    where $n \in [0,N_p-1]$ is the ULA antenna element index, $\varrho$ is the scaling parameter for the phase perturbation, and the parameter $\Pi$ controls the exponent power applied to the position-dependent term. The beam broadening vector $\wv_{\rm PPF}$ has coefficients
$w_{\rm PPF}(n) = \exp(jf(n))$.
In \eqref{eq:ppf}, increasing $\varrho$ results in a broader beam for a fixed $\Pi$, whereas decreasing $\Pi$ widens the beam for a fixed $\varrho$.
The resulting vector of ULA coefficients after phase perturbation is given by 
$\wv_{\rm PPF} \odot \wv_{\rm binary} \odot \qv$. 
For example, the widened flat top beam of Fig.~\ref{fig:step2} is obtained for  $\varrho=2$ and $\Pi=1$. 

\begin{figure}[h!] 
%
%
        \centering
        \includegraphics[width=8cm]{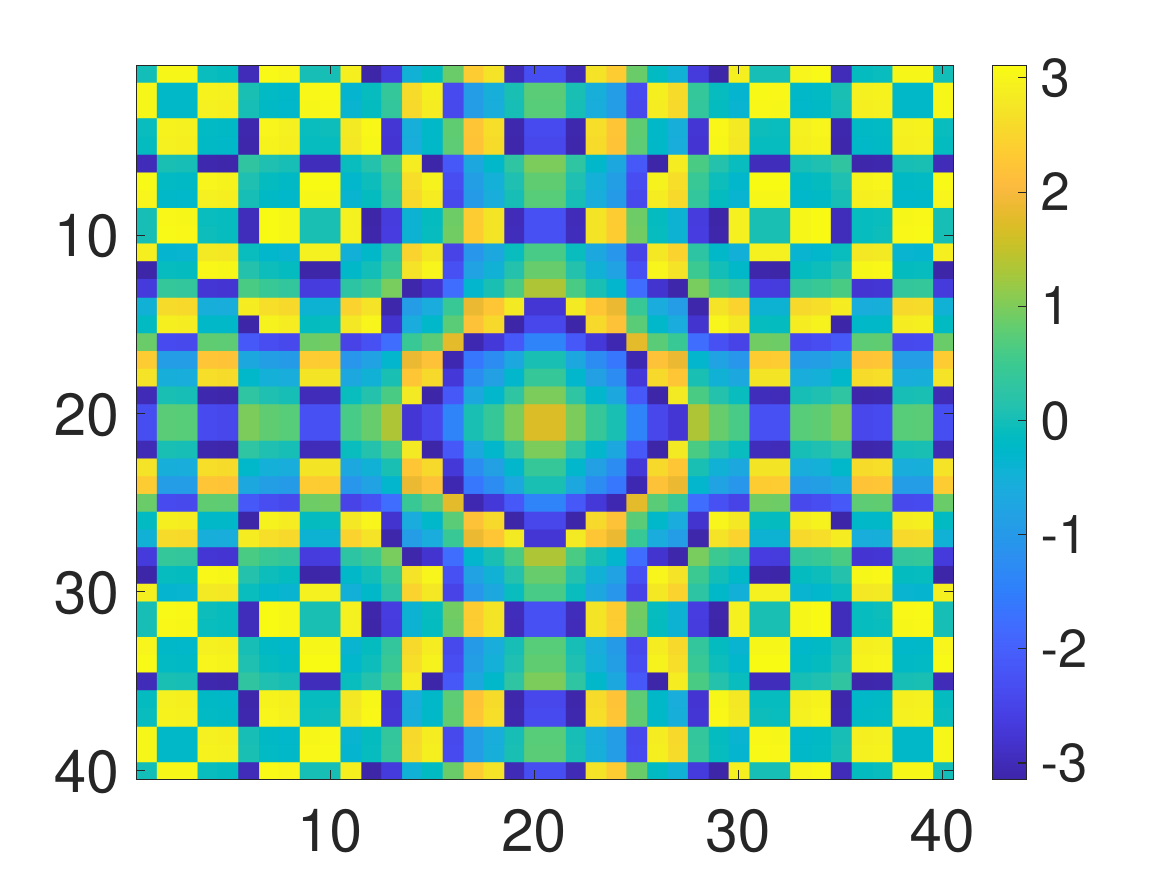}
        \caption{A locally optimum phase profile, corresponding to the wider flat-top beam in Fig. \ref{fig:step4} (applied in both the dimensions).}
        \label{fig:Step4HeatMap}
%
\end{figure} 
   
\subsubsection{Local Optimization}

The pragmatic phase perturbation method described above serves to produce a good starting point $\wv_{\rm ini} = \wv_{\rm PPF} \odot \wv_{\rm binary}$ 
for a successive convex relaxation optimization algorithm adapted from \cite{ghanem2022optimization} to our case (see Appendix \ref{appendix:optimization} for details), yielding the final phase-only vector $\wv_{\rm opt}$ for further improved/controlled beam shapes. 
In particular, the resulting radiation pattern of the ULA with coefficients $\wv_{\rm opt} \odot \qv$ achieves tunable flat-top beam width with small passband ripple, as shown in Fig.~\ref{fig:step4} (to be compared with 
the corresponding results before the optimization in Fig.~\ref{fig:step2}).
The planar RIS phase configuration of $\Wm_{\rm opt} = \wv_{\rm opt} \times \wv_{\rm opt})$ corresponding to the wider flat-top beamshape of Fig.~\ref{fig:step4} is shown in Fig.~\ref{fig:Step4HeatMap}. 
  
\begin{rem} \label{rem2}
The semi-definite relaxation and the first order Taylor approximation approach
of \cite{ghanem2022optimization} yields a local optimum that is strongly dependent on the initialization of the iterative optimization algorithm. In \cite{ghanem2022optimization}, no particular insight is devoted to how to find an efficient initialization. This problem is further exacerbated in our case because of the imposed fixed amplitude taper of $\qv$. We have extensively tested the approach with various initializations and we have verified that our approach of using \eqref{eq:ppf} to create a good initial point for the optimization in \cite{ghanem2022optimization} yields very satisfactory results. 
\hfill $\lozenge$
\end{rem}

\subsection{Extension to Planar Array}
\label{subsec:planar_extension}

\begin{figure*}[h!] \vspace{0cm}
    \centering
    \begin{subfigure}[b]{0.32\textwidth}
        \includegraphics[width=\linewidth]{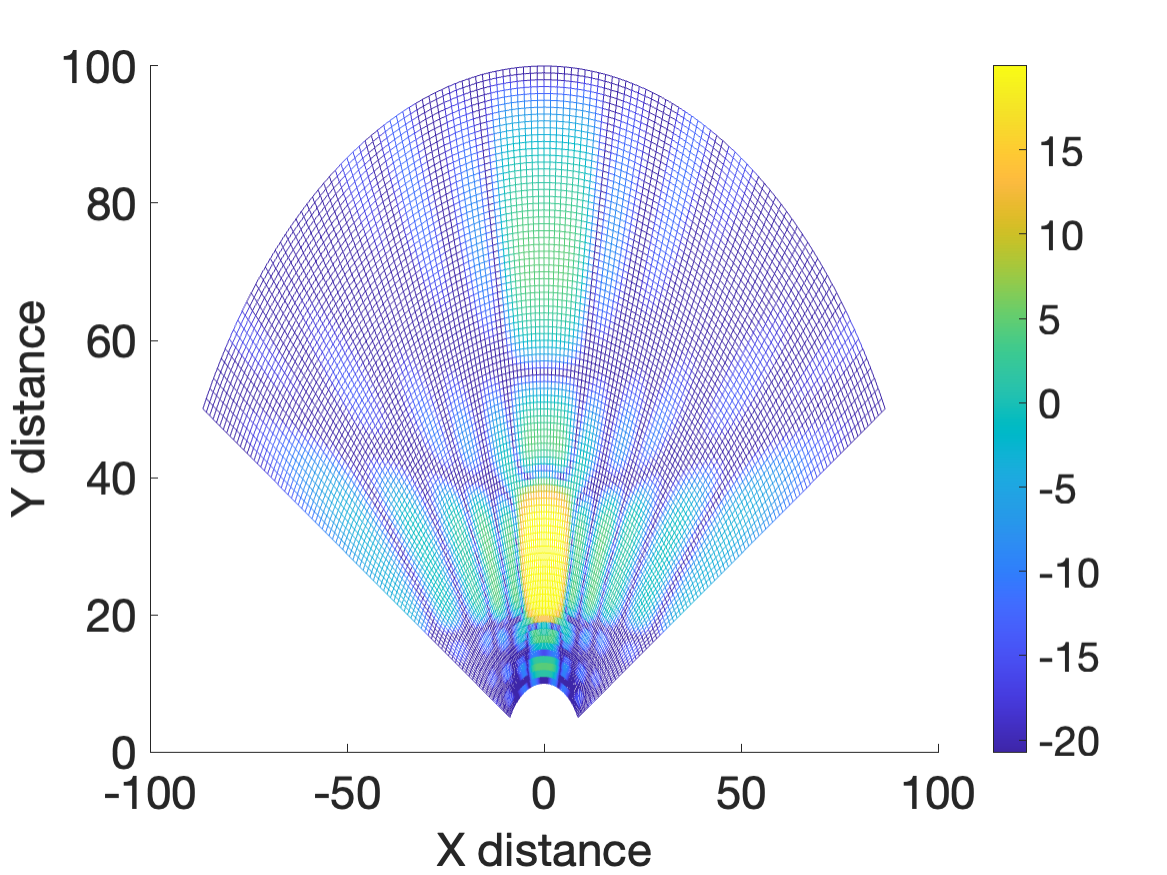}
        \caption{Confined flat top beam from the binary phase profile $\wv_{\rm binary}$.}
        \label{fig:confined_step_2}
    \end{subfigure}
    \hfill
    \begin{subfigure}[b]{0.32\textwidth}
        \includegraphics[width=\linewidth]{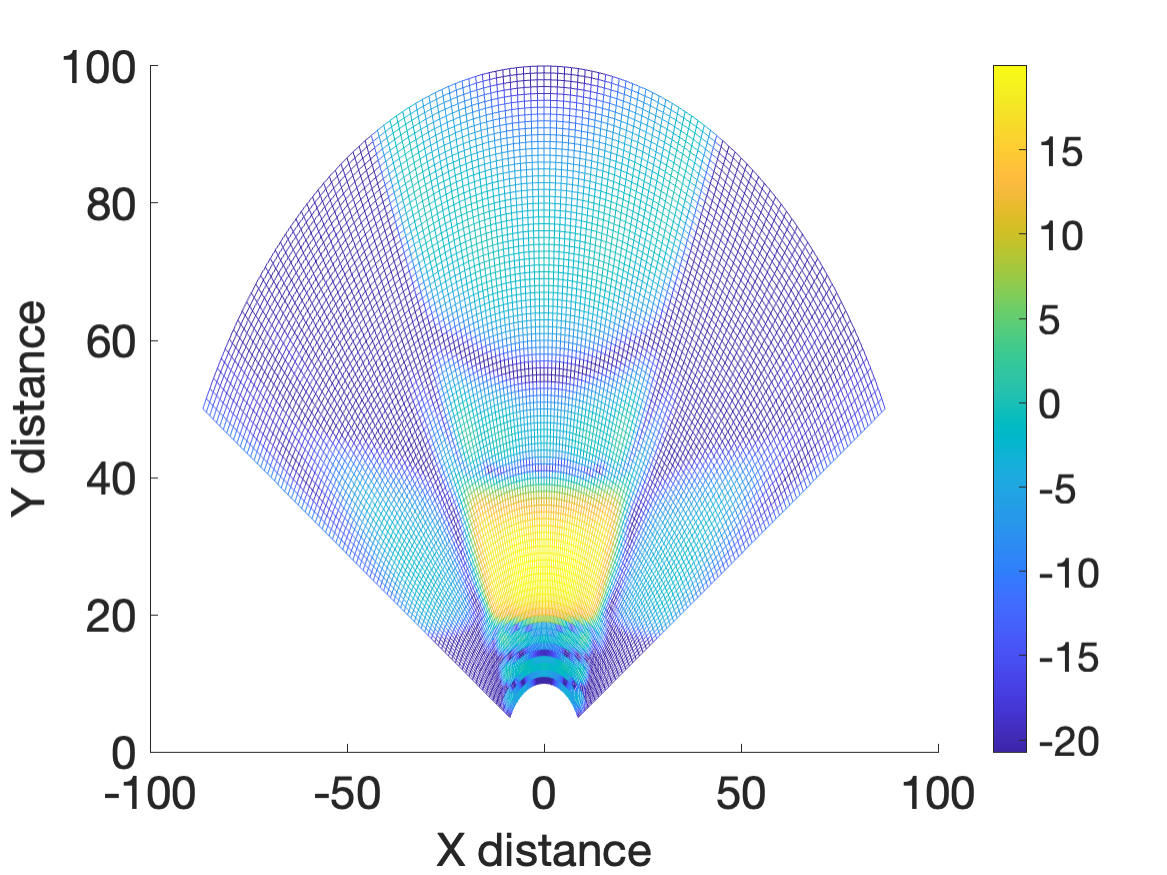}
        \caption{Azimuth widened flat top beam from the local optimization}
        \label{fig:azimuth_step_4}
    \end{subfigure}
    \hfill
    \begin{subfigure}[b]{0.32\textwidth}
        \includegraphics[width=\linewidth]{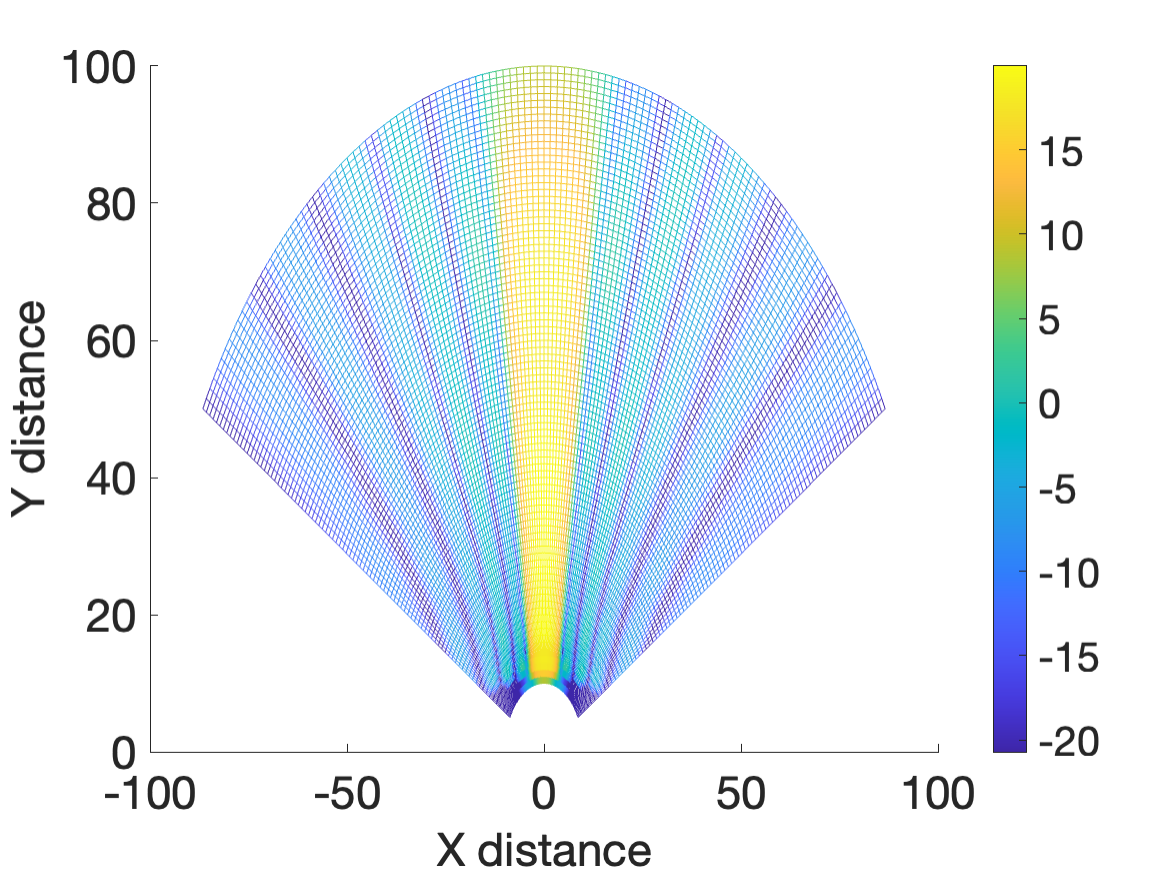}
        \caption{Elevation widened flat top beam from the local optimization}
        \label{fig:elevation_step_4}
    \end{subfigure}
    \caption{Ground footprints of flat-top beams from the binary phase profile and the final local optimization to shape the beamwidth in azimuth and elevation. The wide beam shape of Fig. \ref{fig:step4} is used to widen in the respective azimuth or elevation dimension.}
    \label{fig:gnd_flat_tops}
\vspace{0cm} \end{figure*}

As anticipated before, we shall form the beam shapeing phase control of the RIS elements as $\Wm = \Wm_{\rm opt} = \wv_{\rm opt} \times \wv_{\rm opt})$ as described before. Then, array of complex RIS elements coefficients is obtained as  $\Am(\phi_0,\theta_0; 0) \odot \Wm_{\rm opt} \odot \widetilde{\Wm} \odot |\Um_1|$. Notice that the resulting beam pattern is due to the true (generally non-separable) amplitude profile $|\Um_1|$, despite we have used its separable approximation only for the purpose of reduced complexity design.

Fig.~\ref{fig:gnd_flat_tops} shows representative ground footprints obtained from various optimization steps. In particular, Fig. \ref{fig:confined_step_2} shows the footprint of the boresight direction beam ($\phi_0 = \theta_0 = 0$) for the AMAF$_2$-RIS$_{40}$ without phase perturbation and optimization, i.e., when  $\Wm = \wv_{\text{binary}} \times \wv_{\text{binary}}$. 
Figs. \ref{fig:azimuth_step_4} and  \ref{fig:elevation_step_4} show the case where 
$\Wm = \Wm_{\rm opt} = \wv_{\rm opt}^x \times \wv_{\rm opt}^z$ 
with phase perturbation \eqref{eq:ppf} and local optimization, where the 
$x$-component and the $z$-component (controlling azimuth and elevation, respectively) are optimized using different target beam bandwidths. In 
Fig. \ref{fig:azimuth_step_4} shows a beam with narrow elevation width 
and wide azimuth width. 
Fig. \ref{fig:elevation_step_4} shows 
a beam with wide elevation width 
and narrow azimuth width. This demonstrates the flexibility and effectiveness of our pragmatic beam shaping design approach.

\subsection{Hierarchical Codebook and Effective Channel}
\label{subsec:hierarchical_codebook}

\begin{figure*}[ht!] \vspace{0cm}
    \centering
    \begin{subfigure}[b]{0.32\textwidth}
        \includegraphics[width=\linewidth]{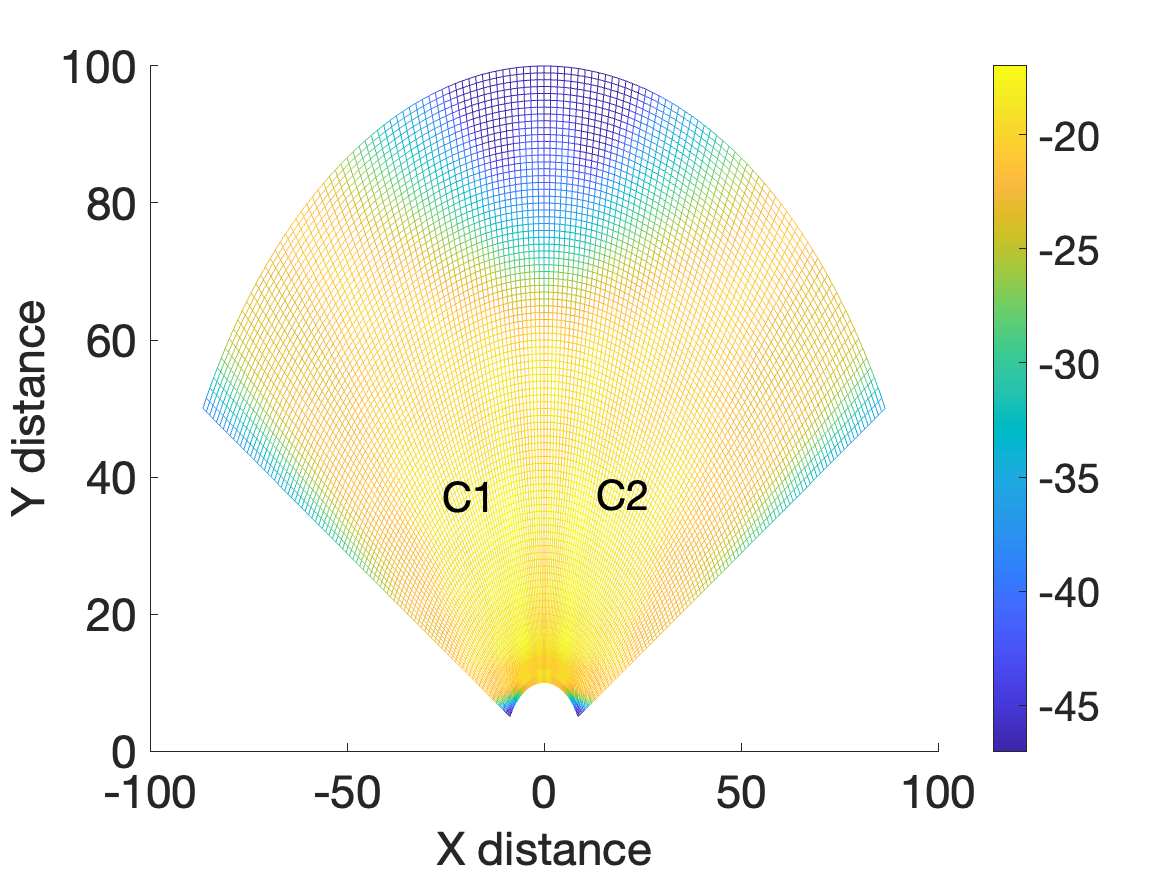}
        \caption{Level 1}
        \label{fig:Level1onGround}
    \end{subfigure}
    \hfill
    \begin{subfigure}[b]{0.32\textwidth}
        \includegraphics[width=\linewidth]{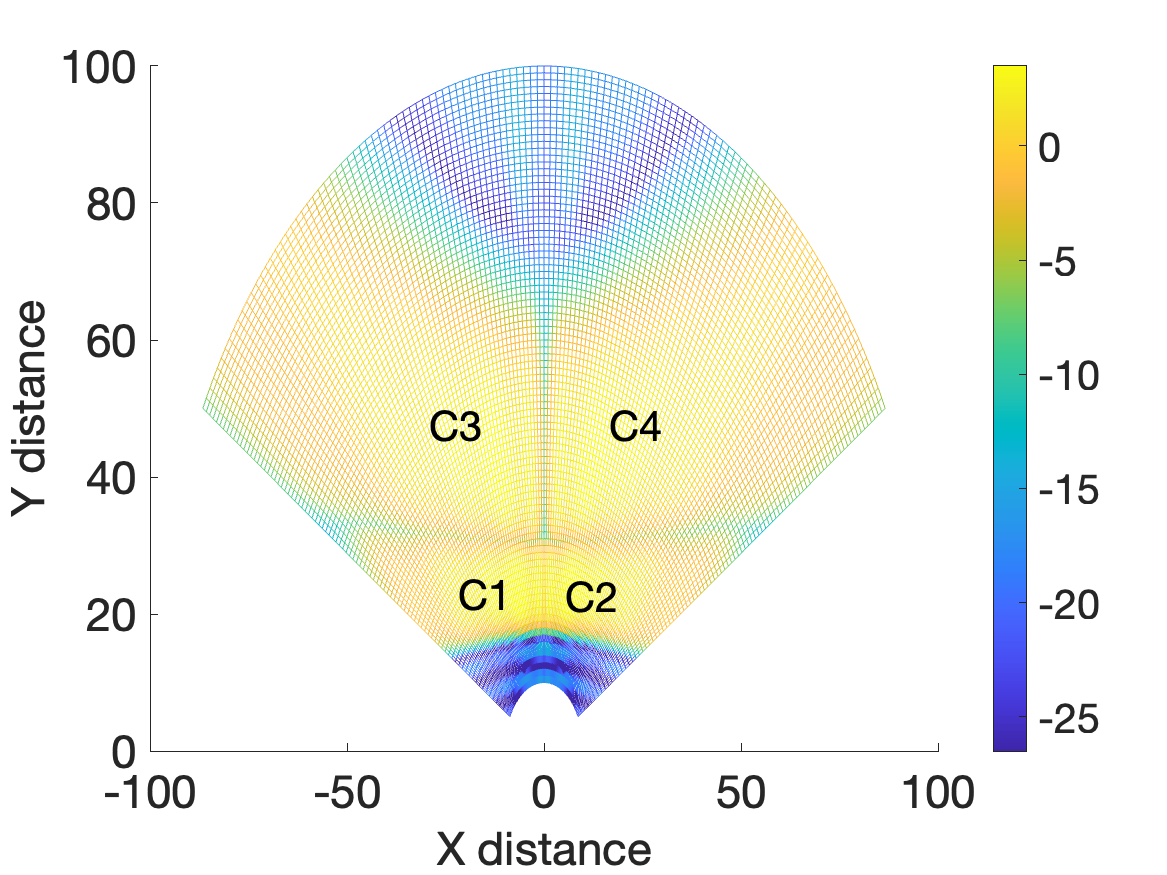}
        \caption{Level 2}
         \label{fig:Level2onGround}
    \end{subfigure}
    \hfill
    \begin{subfigure}[b]{0.32\textwidth}
        \includegraphics[width=\linewidth]{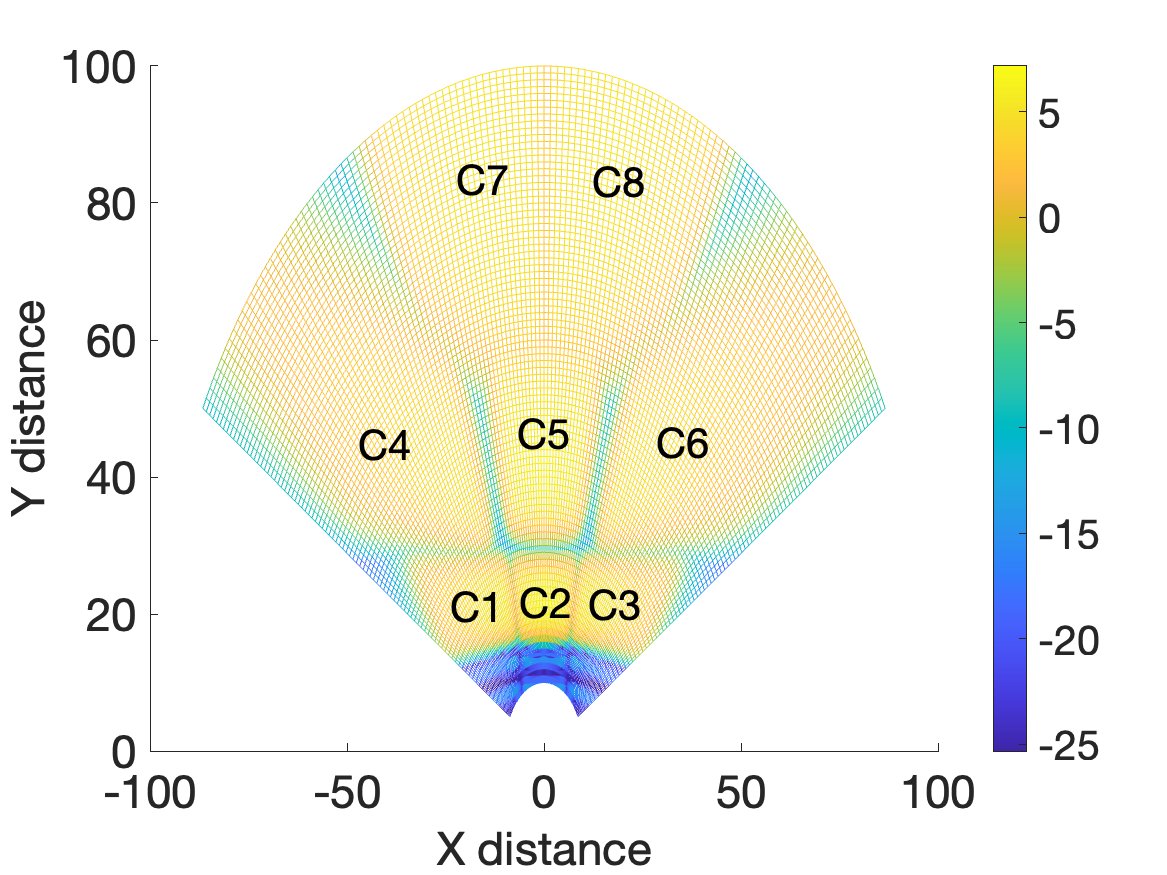}
        \caption{Level 3}
        \label{fig:Level3onGround}
    \end{subfigure}
    \caption{Level 1--3 hierarchical beamforming codebook (beam ground footprints). Level 3 codes C7 and C8 correspond to the case in levels 1 and 2 when the RSRP is very poor at the edge of the cell in the center between the left and the right halves.}
    \label{fig:LevelAll}
\vspace{0cm} \end{figure*}

Using the method described above, we can obtain beamforming codebooks with a good level of flexibility, trading off coverage and beamforming gain. 
As said before, the codebook $\{\Xim^{(c)} : c \in [1:C]\}$ is obtained by combining suitably designed beam shapes (defined by suitably optimized $\Wm$) and center beam 
steering angles determining $\Am(\phi_c,\theta_c;0)$ for angles
$(\phi_c, \theta_c) : c \in [1:C]$.

For example, Fig.~\ref{fig:LevelAll} shows a three-level  hierarchical beamforming codebook
for the AMAF$_2$-RIS$_{40}$ configuration, suitable for successive bisection of the coverage sector. In general, the nested hierarchy allows the application of bisection methods as in \cite{Hybrid_mmWave}. This may lead to energy saving in the case only a few users need to be associated to beams. Otherwise, all beams at level 3 can be searched simultaneously, by transmitting orthogonal ZC sequences through all the $K = 8$ antenna ports. 

The number $C$ of beams at the highest level of the hierarchy needs not be equal to the number $K$ of AMAF-RIS modules. After each UEs $k = 1, \ldots, U$ is associated to its best beam $c_k$ (see Section \ref{subsec:rsrp_computation}), 
the scheduler dynamically selects groups of up to $K$ out of $U$ users 
belonging to $K$ out of $C$ distinct beams to be served using spatial multiplexing. 
By dynamically selecting different groups of users such that each user has the same 
probability of being selected, fairness among the users can be achieved. 
 
 Assume a selected user group $\{k_1,\ldots, k_K\} \subseteq [1:U]$ of size $K$ with distinct associated beam indices, i.e., such that  $c^{k_i} \neq c^{k_j}$ for all $i\neq j$ and, without loss of generality, assume that UE $k_i$ is served by the data stream sent by the $i$-th basic module. The resulting $K \times K$ effective baseband channel matrix on OFDM subcarrier $f_\nu$, denoted by $\Hm(f_\nu)$, has elements (using \eqref{effective channel})
\begin{equation}
[\Hm(f_\nu)]_{i,j} = \sum_{\ell=1}^K \gv_{k_i,\ell}(f_\nu)^\herm \diag(\vec(\Xim^{(c_{k_\ell})}) \Tm_{\ell,j}(f_\nu) \vv_1. \label{effective channel1}
\end{equation}
where $\Xim^{(c_{k_\ell})}$ is the beamforming codeword associated to UE $k_\ell$. 

The effective baseband channel matrix can be estimated 
estimated using standard SRS/DMRS piloting, fully compatible with the 5GNR standard \cite{Book5GNR, SRSref, 38.211}. Then, we shall consider the effect of digital baseband ZF precoding applied to the effective channel $\Hm(f_\nu)$ on a per-subcarier basis. 

\begin{figure}[h!] 
\centerline{\includegraphics[width=7.5cm]{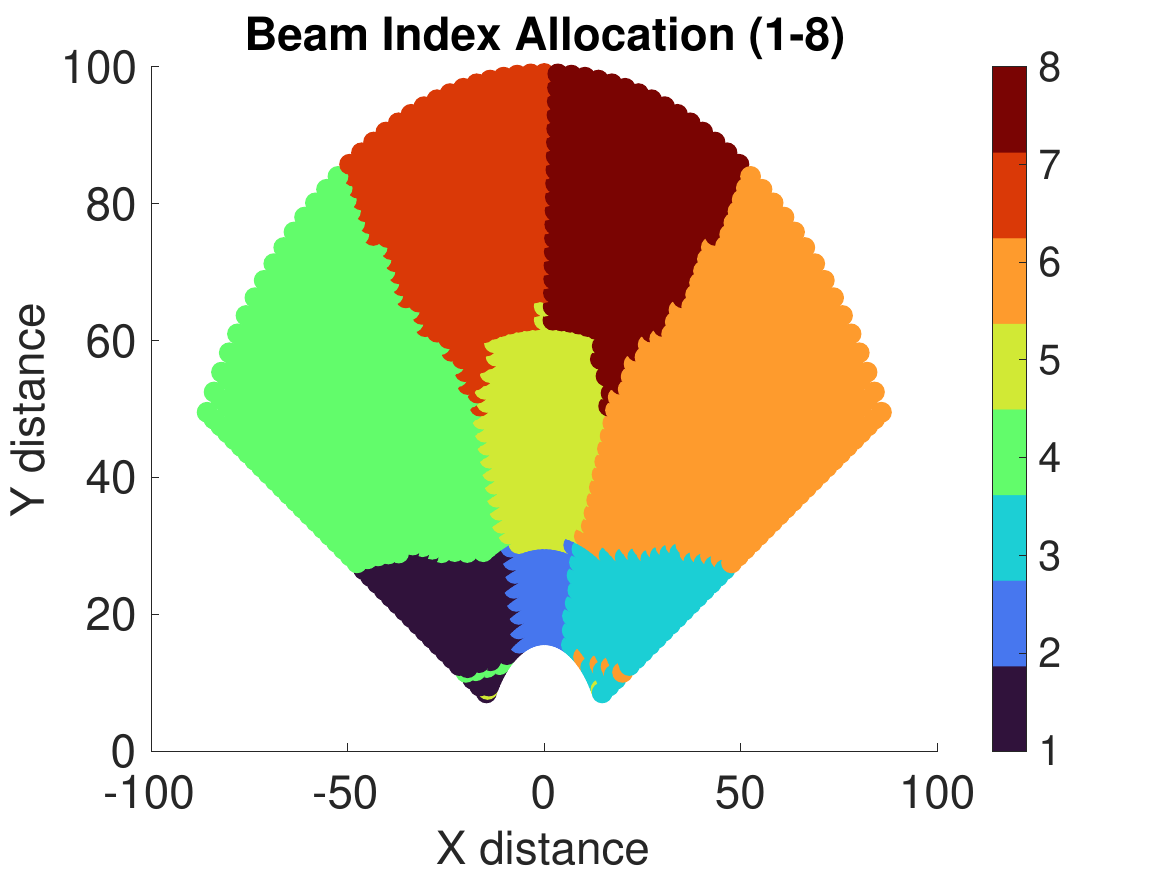}}
\caption{Ground pixelwise best beam selections under pure LOS propagation.}
\label{fig:PureLOSNoScatterer_Beam_Selections}
\end{figure}

In Fig. \ref{fig:PureLOSNoScatterer_Beam_Selections}, we show the best beam selection 
(beam indices are color-coded) for all positions on the coverage sector 
under pure LOS propagation. All UEs located on pixels with the same color, are associated to the same beam. We see than in the case of pure LOS propagation, these areas correspond to the beam footprints in Fig.~\ref{fig:Level3onGround}. In the next section, we shall see that this is no longer the case in the presence of multipath propagation.  

\section{System Set-up and Performance Evaluation}
\label{sec:system}

\begin{table}[h!]
\caption{System requirement specifications.}
\label{tab:SRS}
\centering
\setlength{\tabcolsep}{2pt} 
\begin{tabular}{lclc}
\hline\hline 
Specification & Value & Specification & Value \\
\hline 
Carrier freq. (GHz)       & 100 & Receive noise pow. (dBm)     & -72   \\
Cell range (m)            & 17 to 100  & Receive SNR (dB)  & 0 \\
Azimuth span ($\phi$)     & +/-60$\degree$ & Receive signal power (dBm) & -72 \\
Bandwidth $W$ (GHz) & 5  &  Path Loss $L^{\text{LOS}}_{\rm max}$ (dB) & 112.7  \\
Thermal noise pow. (dBm)  & -77  & EIRP $P_{\rm EIRP}$ (dBm)    & 40.7  \\
Rx NF (dB)      & 5  & RIS size ($N_p\times N_p$)  & 40 x 40  \\
\hline 
\end{tabular}
\end{table}

We consider a system operating in the high mmWave frequencies with $W = 5$ GHz of signal bandwidth at carrier $f_0 = 100$ GHz. The system specifications are summarized in Table \ref{tab:SRS}. 
The BS is formed by $K = 8$ AMAF$_2$-RIS$_{40}$ basic modules, using the codebook with beam footprints in Fig.~\ref{fig:Level3onGround}. The
RF power at each AMAF antenna port is set to $P_{\rm RF} = 32~\mathrm{dBm}$ in order to achieve 0 dB of receiver SNR for an isotropic receiver placed at the sector edge.

\subsection{3D Multipath Scenarios and Channel Characterization}
\label{subsec:3d_multipath_setup}

\begin{figure*}[ht!]
\centering
\begin{tabular}{cc}
\includegraphics[width=0.48\linewidth]{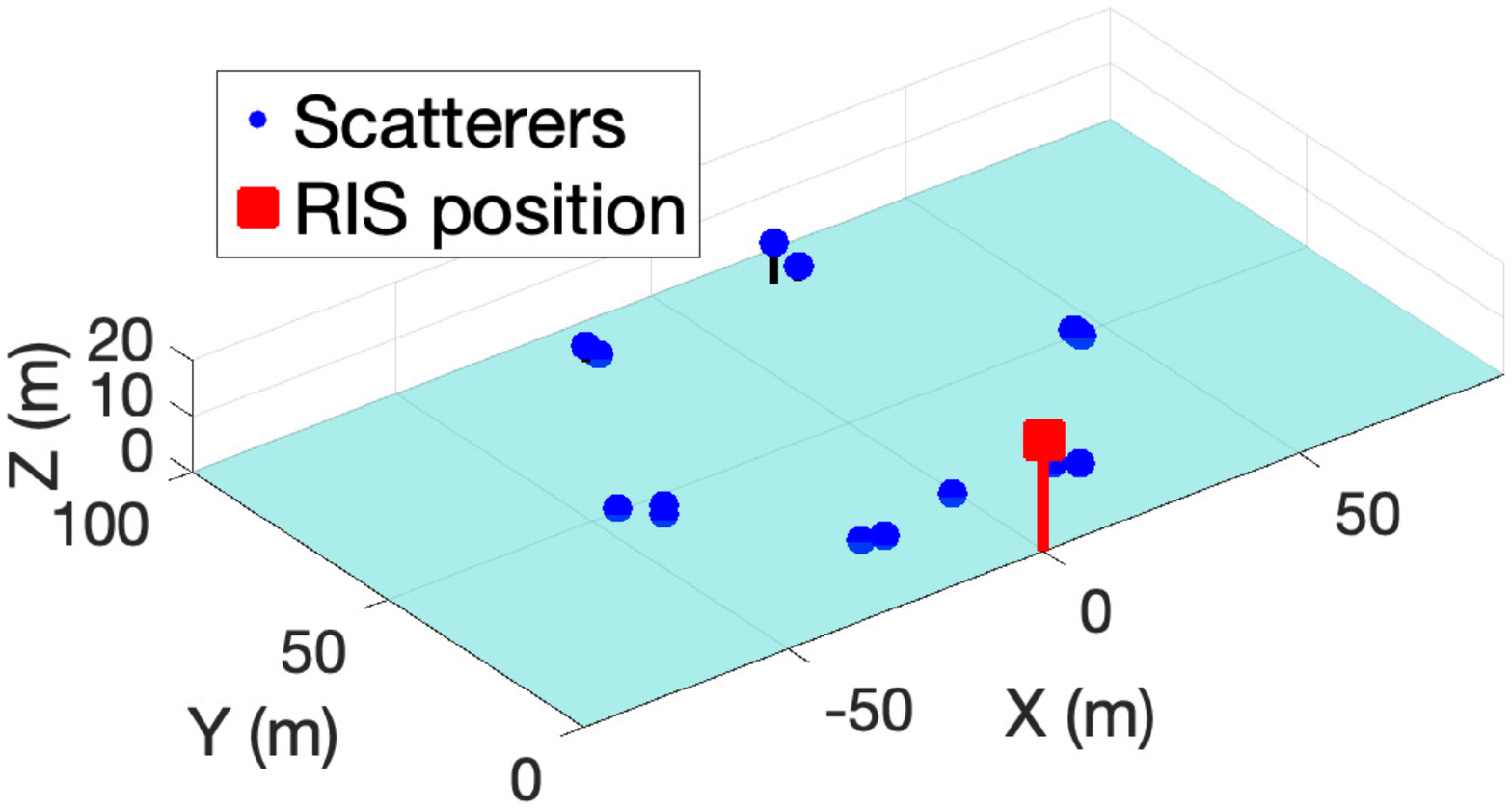} &
\includegraphics[width=0.45\linewidth]{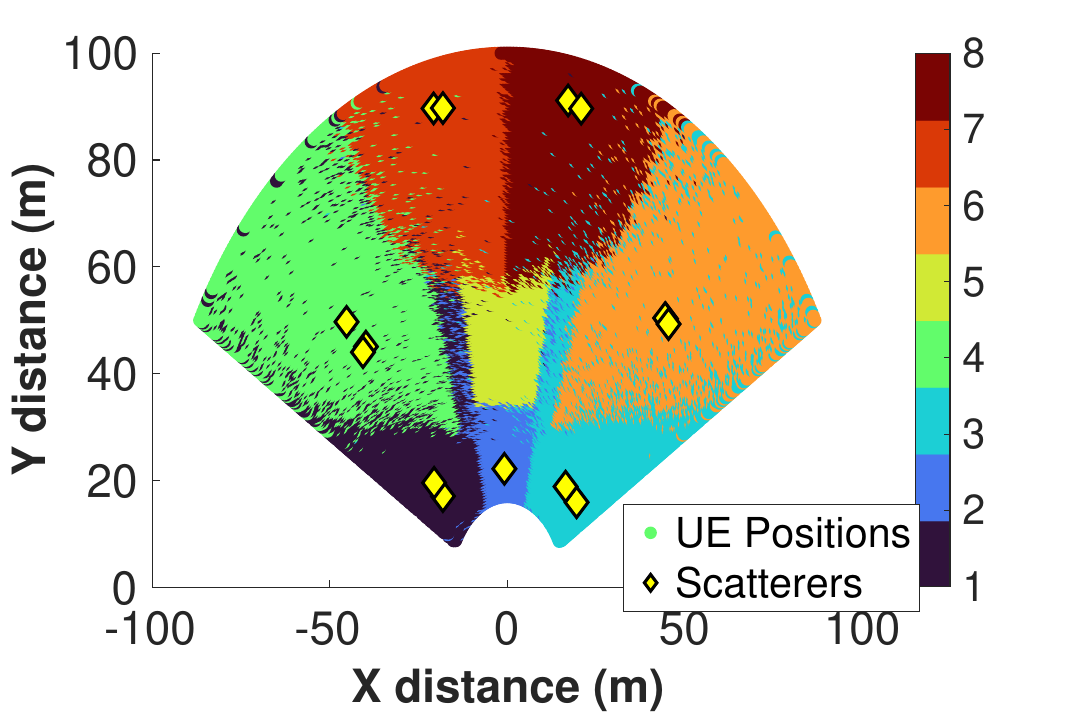} \\
(a) Scenario 1: 15 scatterers from 7 vMF clusters & (b) Scenario 1: Beam selection map \\[0.05em]
\includegraphics[width=0.48\linewidth]{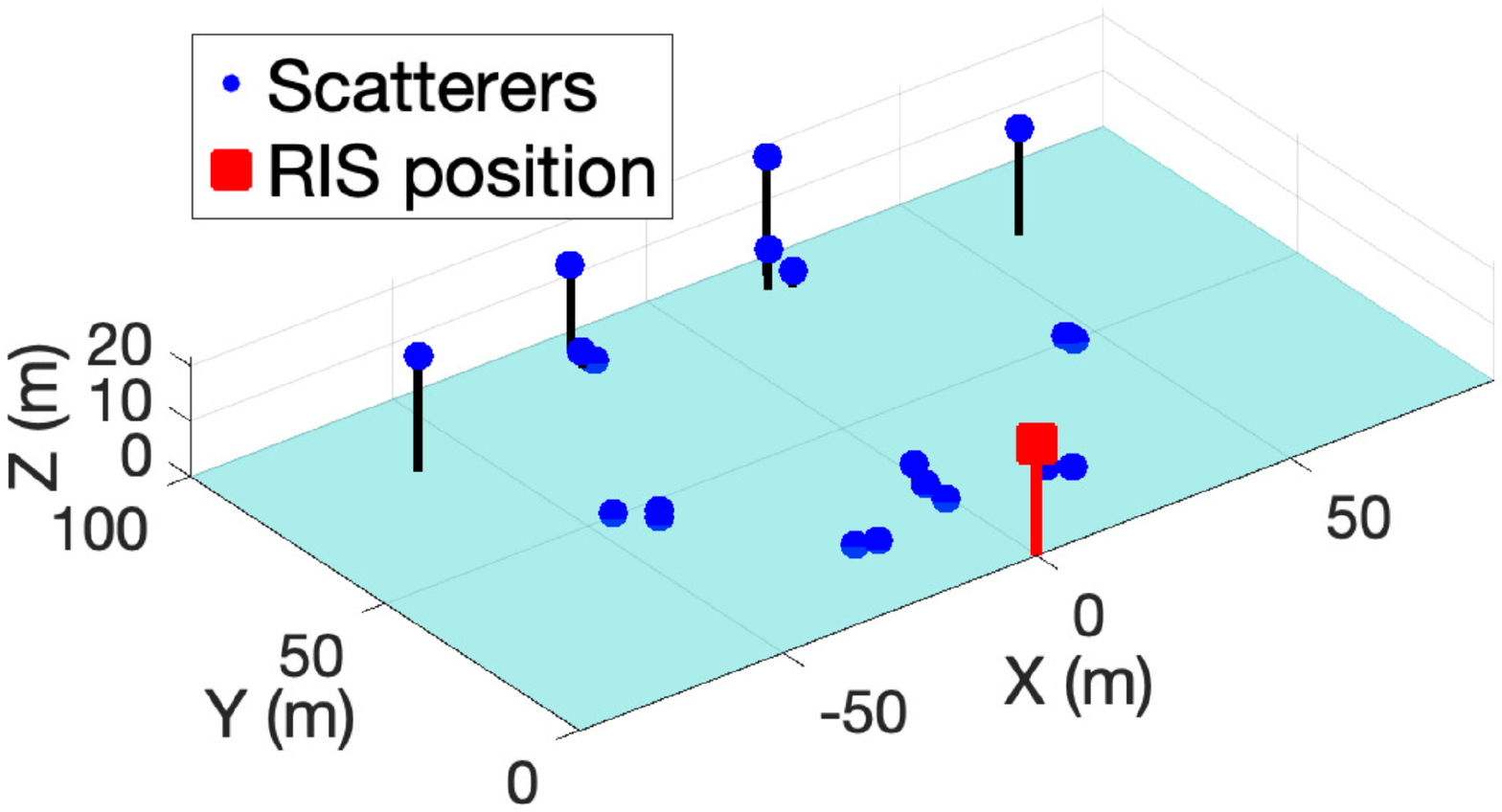} &
\includegraphics[width=0.45\linewidth]{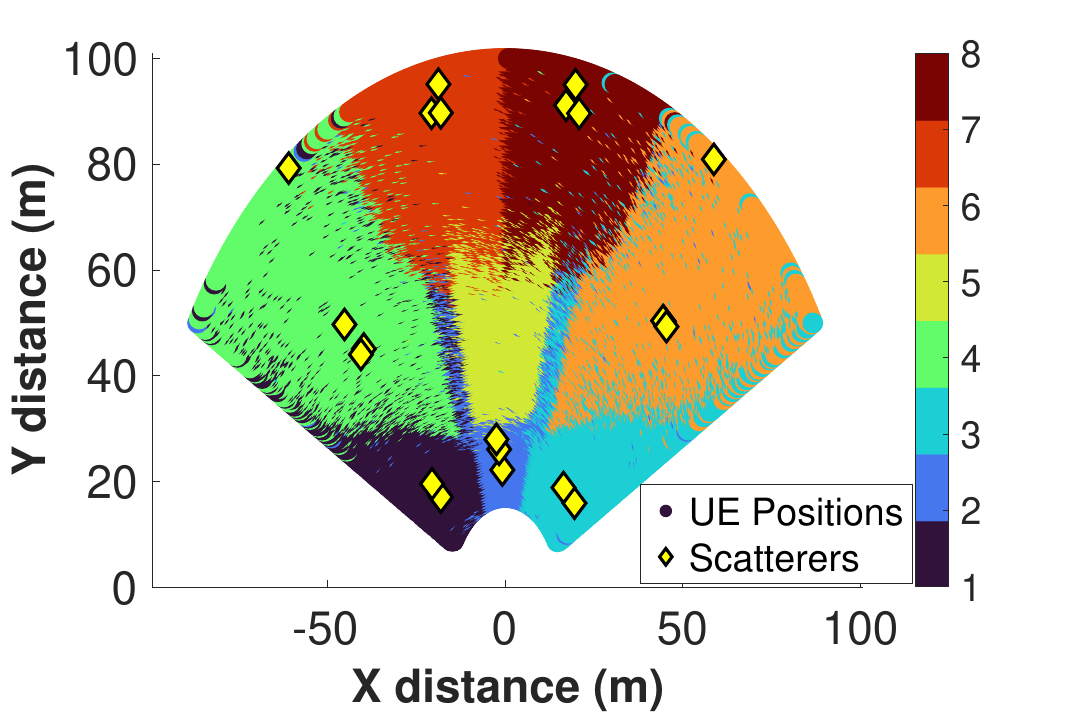} \\
(c) Scenario 2: 21 scatterers from 11 vMF clusters & (d) Scenario 2: Beam selection map \\
\end{tabular}
\caption{3D multipath scenario characterization: (a,b) Scenario 1 - suburban environment with 15 scatterers; (c,d) Scenario 2 - urban environment with 21 scatterers including elevated structures. The beam selection maps illustrate how multipath scattering influences beam assignments across the coverage area.}
\label{fig:multipath_scenarios}
\end{figure*}

We consider two different 3D multipath scenarios with geometrically consistent scattering distributions, corresponding to different parameters in the vMF distributed scatterer model introduced in Section~\ref{subsec:multipath_model}.

\subsubsection{Scenario 1: Suburban Environment}
\label{subsubsec:scenario1}

The first scenario models a suburban environment with 15 scatterers distributed across 7 clusters, as illustrated in Fig.~\ref{fig:multipath_scenarios}. The concentration parameter $\kappa_c$ for each cluster follows an inverse-square relationship with distance
$\kappa_c \propto \frac{1}{\rho^2}$, 
where $\rho$ represents the scatterer range from the BS. This is physically motivated by the reduced solid angle subtended by objects of fixed size at greater distances. 

We establish a reference concentration $\kappa_0 = 390$ at the RIS boresight ground interception point (ground range 26.19 m, azimuth $0^\circ$), corresponding to scatterers of approximately 1.5 m height typical of street furniture. This scaling yields concentration factors ranging up to $\kappa_{\text{max}} = 4 \cdot 10^3$ towards the cell edge. We sample $[2, 1, 2, 3, 3, 2, 2]$ elements per cluster, totaling 15 scatterers that represent pedestrians, vehicles, and other common obstructions.

\subsubsection{Scenario 2: Urban with Elevated Structures}
\label{subsubsec:scenario2}

The second scenario models urban environment with additional elevated structures, comprising 21 scatterers across 11 clusters (Fig.~\ref{fig:multipath_scenarios}). This configuration introduces four additional clusters at $\approx 20$ m height, representing billboards, building elements, or other elevated urban infrastructure at cell edges.
Two supplementary scatterers are added to the the central cluster to model increased local scattering near the BS. Also here we use the $\kappa_c \propto 1/\rho^2$ scaling for physical consistency.

\subsection{Performance Evaluation}
\label{subsec:performance_methodology}

We use Monte Carlo simulation as follows:
\begin{itemize}
\item 100 independent user drops with $U = 64$ users uniformly distributed across the coverage sector (azimuth: $\pm60^\circ$, ground range: 17--100 m);
\item Beam-wise user partitioning where each user reports its best-serving beam according to the RSRP metric, creating beam-specific user pools of users;
\item Fair airtime allocation among users by running a large number of user group selections by selecting at random with uniform probability one user per beam-specific pool.
\end{itemize}
In particular, we are interested in assessing the benefits of baseband ZF precoding
applied to the resulting effective channels as defined in 
\eqref{effective channel}. Therefore, we consider the performance with pure RF beamforming 
as in \cite{NewOldIdea} as a baseline, and compare with the performance obtained 
using ZF with per-antenna port power constraint. Notice that 
since each AMAF-RIS module has its own power amplifier, the meaningful constraint for the baseband precoder is not ``sum power'' (leading to a precoder matrix trace constraint) but per per-antenna port constraint, following the algorithm of \cite{ShamaiZF} (not included in this paper for lack of space).

Consider a given scheduling slot with selected multiuser group $\{k_1, \ldots, k_K\} \subseteq [1:U]$. Let $\Hm(f_\nu)$ denote the resulting 
  effective channel matrix as defined in \eqref{effective channel1}, 
  $P_{\rm RF}$ is AMAF output RF power, 
$N_0$ is the complex baseband AWGN power spectral density, and $W$ is the signal bandwidth.
Also, let $\Gm(f_\nu)$ denote the ZF precoding matrix, calculated for each OFDM subcarrier. Then, the information rate (in bit/s/Hz) supported by user $k_i$ during the slot is given by 
    \begin{equation}
        R_{k_i} = \frac{1}{N_{\text{sub}}} \sum_{\nu=1}^{N_{\text{sub}}} \log_2\left(1 + \text{SINR}_{k_i}[\nu] \right) \quad \text{bits/s/Hz}, \label{ziorate}
    \end{equation}
where, in the case of pure RF beamforming,     
\begin{align}
    \text{SINR}_{k_i}[\nu] & = 
    \frac{| [\Hm(f_\nu)]_{i,i}|^2 P_{\rm RF}}{W N_0 + \sum_{j \neq i}|[\Hm(f_\nu)]_{i,j}|^2 P_{\rm RF}},  
    \label{eq:SINR}
\end{align}
and, in the case of ZF precoding,
\begin{align}
    \text{SINR}_{k_i}[\nu] & = 
    \frac{| [\Hm(f_\nu) \Gm(f_\nu)]_{i,i}|^2 P_{\rm RF}}{W N_0 + \sum_{j \neq i} |[\Hm(f_\nu) \Gm(f_\nu)]_{i,j}|^2 P_{\rm RF}}.
    \label{eq:SINR-ZF}
\end{align}
Notice that when the ZF precoder is calculated on the estimated (e.g., from SRS pilots) 
effective channel matrix, the interference is generally not exactly zero, and therefore we still have a multiuser interference terms in the denominator of \eqref{eq:SINR-ZF}.
    
System performance is quantified in terms of:
\begin{itemize}
    \item Per-Beam Average Spectral Efficiency: for each beam $c$, we collect the per-user information rates
    \eqref{ziorate} over a large number of scheduling slots of the selected users associated to beam $c$ and compute the average information rate. 
    \item Per-beam information rate Cumulative Distribution Functions (CDFs): for the same 
    Monte Carlo generated data set, we present the corresponding CDFs for each beam $c$. 
\end{itemize}

\section{Results and Comparisons} \label{sec:results}


\begin{figure}[ht!]
\centering
\includegraphics[width=\linewidth]{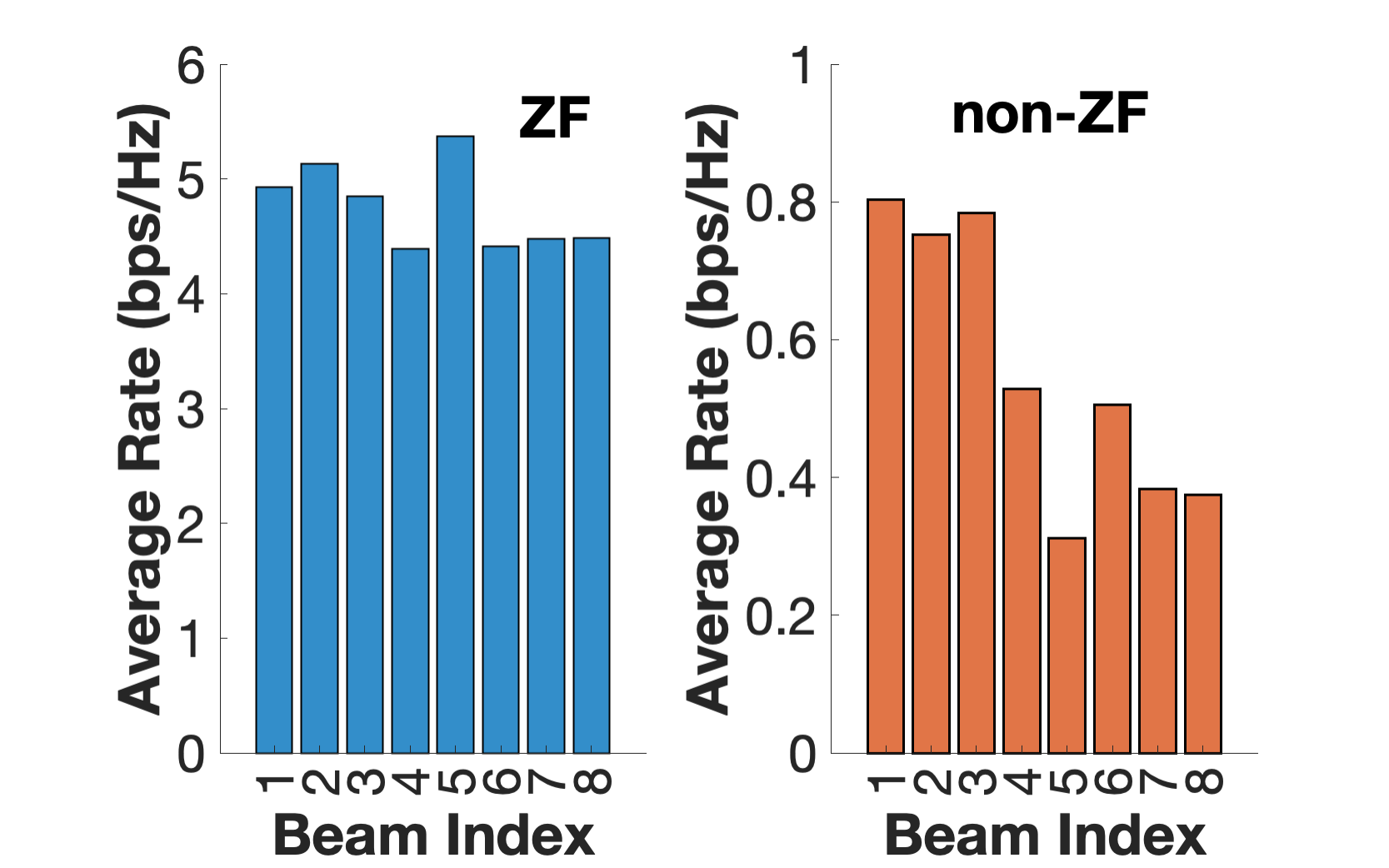}
\caption{Scenario 1: Average beam rates for the baseband ZF precoding and no baseband precoding.}
\label{fig:BeamRates_Scenario1}
\end{figure}

\begin{figure}[ht!]
\centering
\includegraphics[width=\linewidth]{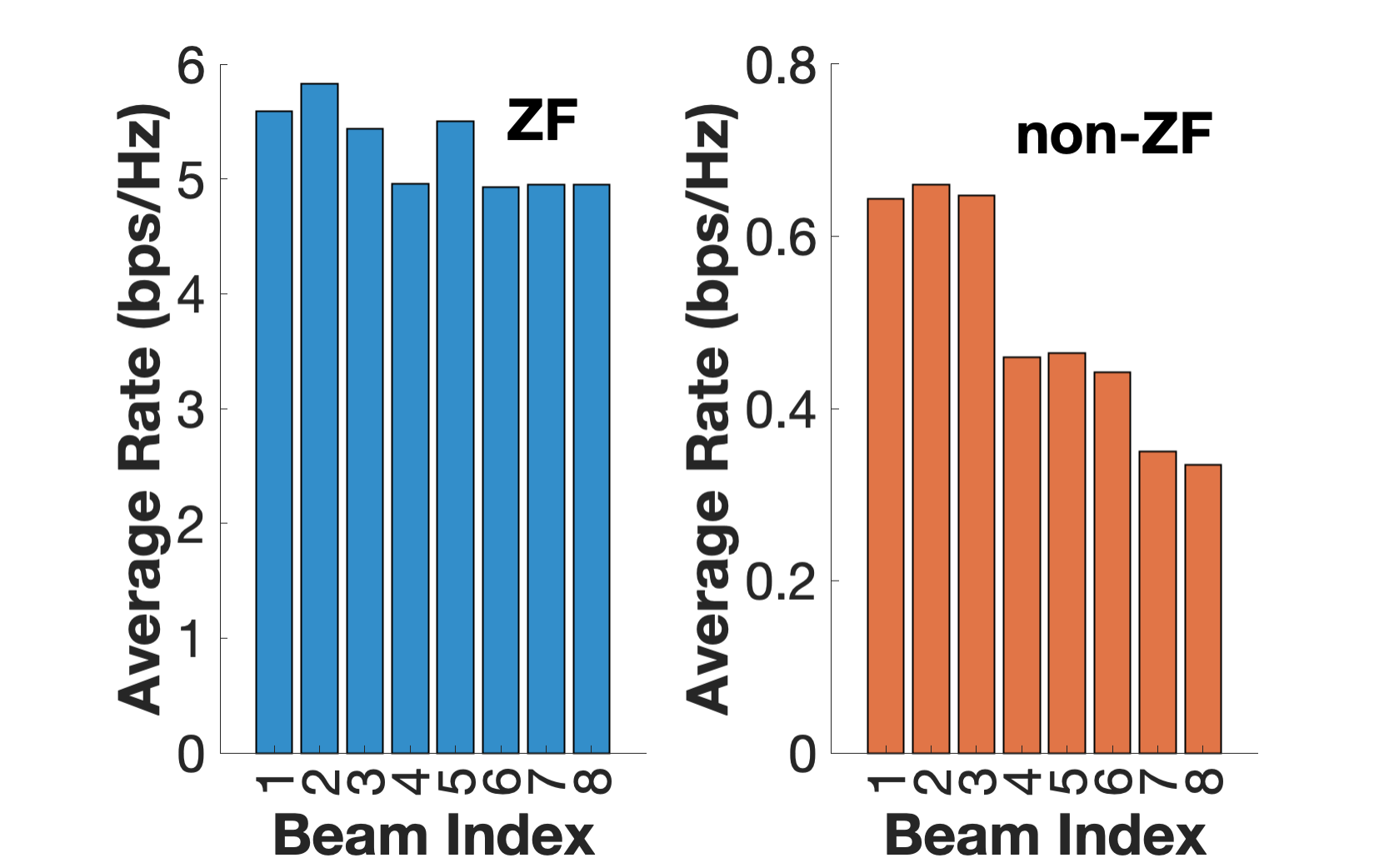}
\caption{Scenario 2: Average beam rates for the baseband ZF precoding and no baseband precoding.}
\label{fig:BeamRates_Scenario2}
\end{figure}

\begin{figure*}[ht!]
\centering
\begin{tabular}{cc}
\includegraphics[width=8cm,height=5cm]{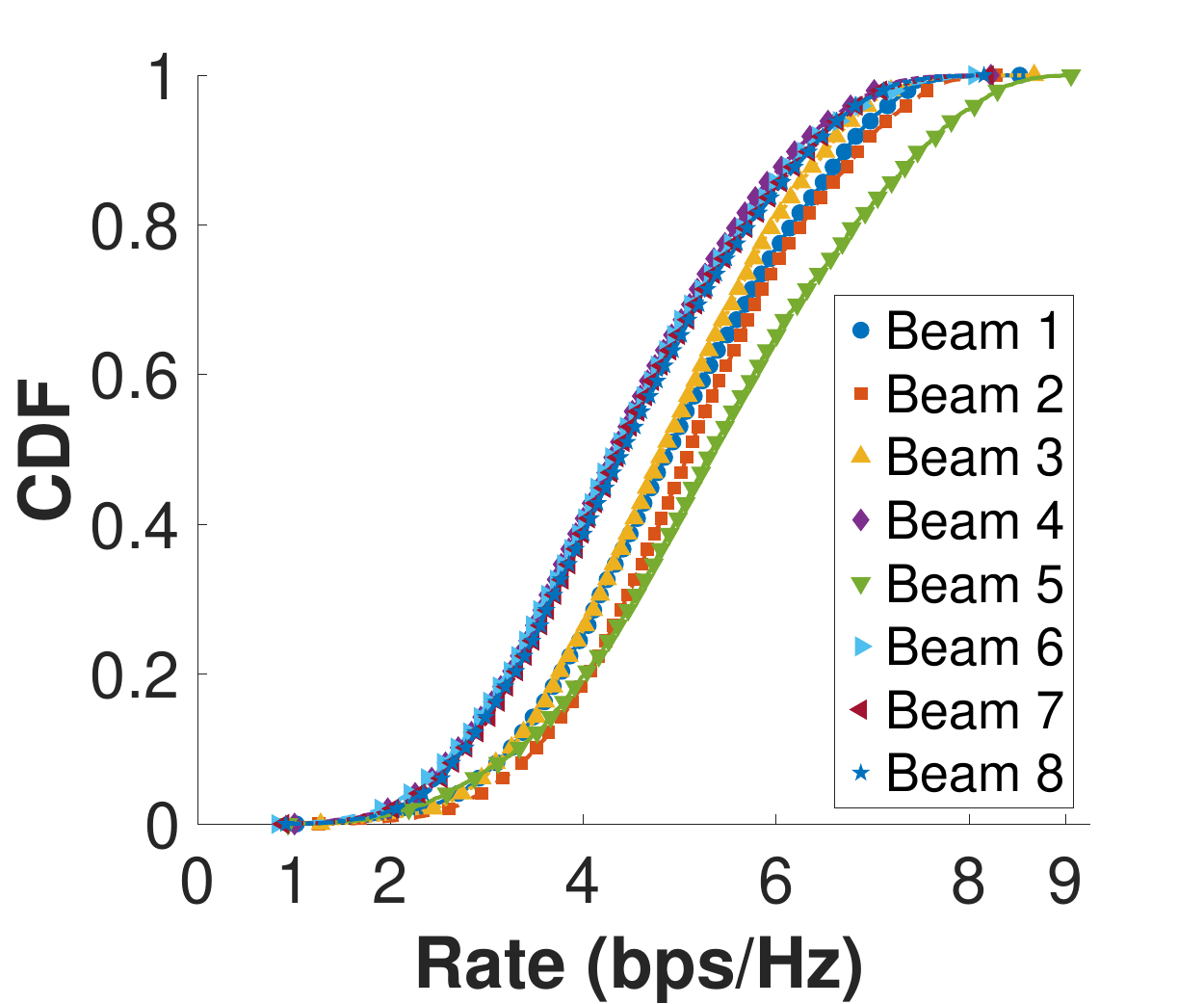} &
\includegraphics[width=8cm,height=5cm]{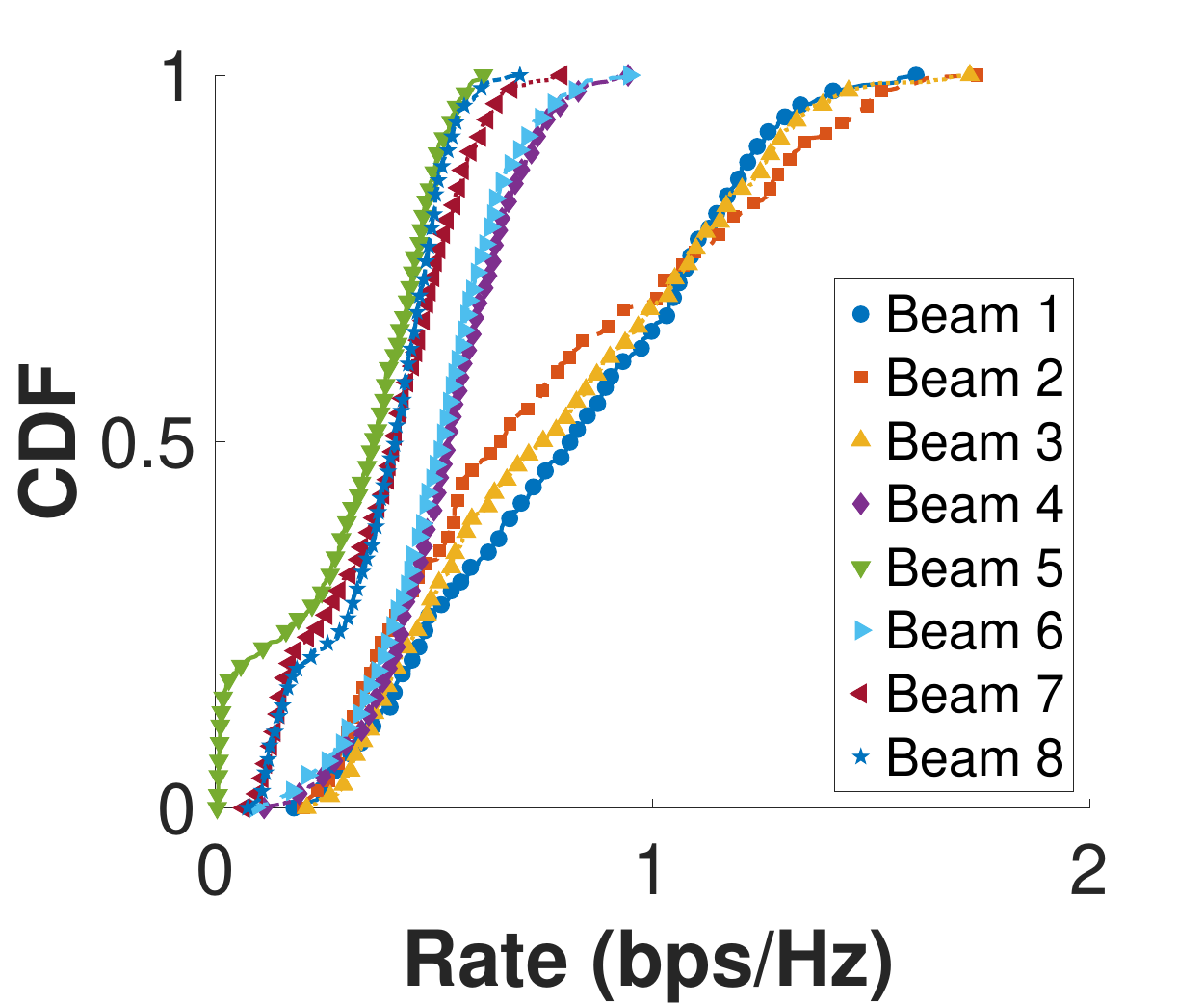} \\
(a) Scenario 1: ZF & (b) Scenario 1: Non-ZF \\[0.05em]
\includegraphics[width=8cm,height=5cm]{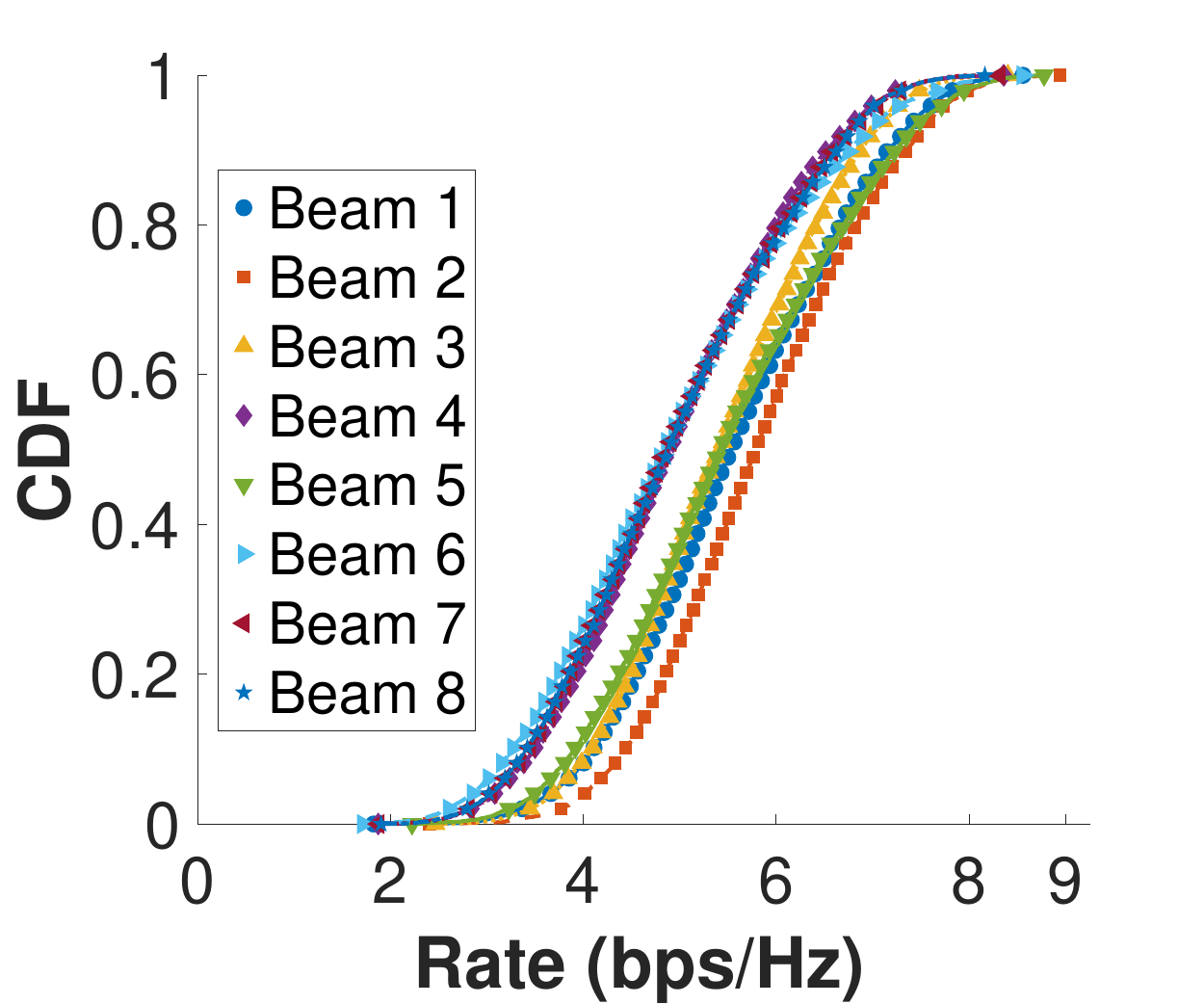} &
\includegraphics[width=8cm,height=5cm]{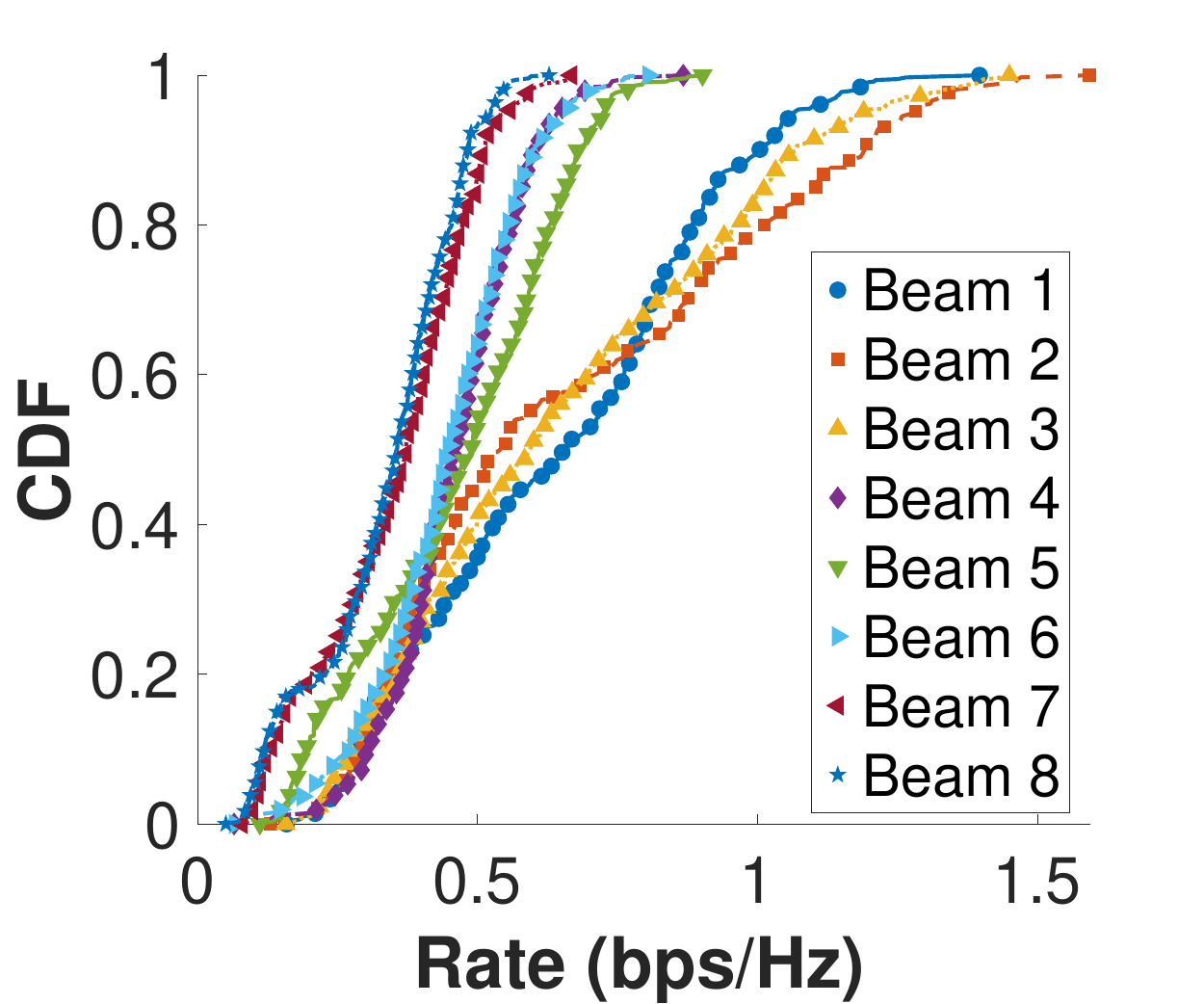} \\
(c) Scenario 2: ZF & (d) Scenario 2: Non-ZF \\
\end{tabular}
\caption{Cumulative distribution functions of user rates per beam for both precoding schemes and scattering scenarios. The ZF precoding demonstrates significantly tighter distributions and improved fairness across all beams, particularly in the challenging multipath conditions of Scenario~2.}
\label{fig:BeamwiseCDFs}
\end{figure*}

The beamwise performance analysis, illustrated in Figs. \ref{fig:BeamRates_Scenario1},  \ref{fig:BeamRates_Scenario2}, and \ref{fig:BeamwiseCDFs}, reveals several critical insights:
\begin{enumerate}[i.]
    \item The baseband ZF precoding successfully recovers near-interference-free performance levels, effectively suppressing the far-end cross-talk (FEXT) introduced by multipath propagation. The substantial performance gap between ZF and non-ZF schemes demonstrates that pure RF beamforming is inadequate in realistic scattering environments, necessitating advanced baseband processing for interference management. 
    The system is interference-limited, meaning the non-ZF performance plateaus at very low rates with increasing transmit power while the ZF performance continues to improve.
    
    \item \textit{Uniform Coverage Performance:} All beams maintain comparable average spectral efficiency under zero-forcing precoding, with variations across beams limited to less than 0.97~bits/s/Hz in Scenario~1 and 0.88~bits/s/Hz in Scenario~2. This consistency demonstrates the flat-top beam codebook's effectiveness in providing uniform coverage throughout the cell sector, despite several challenging factors: the inherent distance-dependent path loss variation across the 17--100~m cell range, the 7~dB element pattern variation $E(\phi,\theta) = 4(\cos\phi\cos\theta)^2$ across the $\pm 60^\circ$ azimuth span, and the concavity of the logarithmic rate function which, by Jensen's inequality, tends to compress rate variations in the high-SNR regime. The beamwise CDFs of user rates in Fig.~\ref{fig:BeamwiseCDFs} further substantiate this quite nicely uniform performance distribution.
    
    \item \textit{Scenario-Dependent ZF Gains:} The gain of baseband ZF precoding over the non-ZF baseline counterpart is slightly marked when the NLOS effects are stronger and exhibits some geometric relations. In Scenario~1, the maximum ZF gain occurs for beam~5 (5.1~bits/s/Hz), which corresponds to the cell-center RIS boresight beam with no scatterers, and the ZF nullifies the interferences from all the scatterers in other beams only. For scenario 2, the maximum ZF gain is for beam 2 ((5.2~bits/s/Hz)) due to the additional scatterers. 
    
\end{enumerate}

\section{Conclusions}  
\label{sec:CONC}

This paper expands upon the AMAF-RIS architecture by extending it from line-of-sight (LOS) to realistic multipath environments through three contributions. First, we present a computationally efficient method for designing phase-only flat-top beams that overcome fixed PEM amplitude constraints. This method produces wide beams with  uniform gain, steep transitions, and low sidelobes, which are ideal for hierarchical beamforming codebooks and have energy efficiency that rivals that of active arrays. 
Second, we used a geometrically consistent 3D multipath channel model using the vMF distribution, enabling realistic simulation of non-isotropic scatterer clusters. The two carefully characterized scenarios, suburban and dense urban with elevated structures, provide a rigorous system performance evaluation under controlled yet representative propagation conditions. Third, a system-level performance evaluation reveals that combining hierarchical codebooks with HDA precoding achieves near-interference-free performance in a multipath MU-MIMO environment. While pure RF beamforming does not perform well in the presence of multipath, analog beam selection combined with the subcarrier-wise baseband ZF precoding provides near-uniform coverage and high average per-beam spectral efficiency. The architecture thus offers a practical, hardware-efficient pathway for RIS-enabled 6G systems at mmWave and sub-THz frequencies.


\appendices

\section{Local Optimization for \\Improved Flat-Top Beam Shaping }
\label{appendix:optimization}

We adapt \cite[Algorithm 1]{ghanem2022optimization} to obtain phase-only RIS coefficients for improved flat-top beam shaping under the fixed nonuniform amplitude taper $\qv$ 
imposed by the PEM design of the AMAF-RIS architecture. 
For a linear RIS with $N_p$ elements, we aim to maximize the minimum gain over angular interval $\mathcal{B} = [\beta_{\min}, \beta_{\max}]$ , where $\beta \triangleq \sin\psi$ (as in \eqref{ULAsteering}).
Let $\wv \in \mathbb{C}^{N_p}$ denote the RIS phase-only vector with $|w_n| = 1$ and define the array steering vector $\av(\beta) = [1, e^{-jkd\beta}, \ldots, e^{-jkd\beta(N_p-1)}]^\transp$, where $k = 2\pi/\lambda_0$ and $d$ is the element spacing. 
The beam pattern accounting for both the fixed amplitude taper and the element radiation pattern is:
\begin{equation}
G(\beta) = E(\beta)|\av(\beta)^\herm (\wv \odot \qv)|^2,
\end{equation}
where $E(\beta)$ is the element factor.

Following the successive convex approximation (SCA) approach in \cite{ghanem2022optimization}, we formulate the optimization problem for a target angular set $\hat{\mathcal{B}}$ (discretized version of $\mathcal{B}$) as:
\begin{subequations} \label{flatopt}
\begin{align}
\underset{\Wm,\alpha}{{\rm maximize}} &\quad \alpha - \eta^{(i)}\Big(\|\Wm\|_* - \|\Wm^{(i)}\|_2 \nonumber\\
&\qquad - \mathrm{Tr}\big[\ev_{\max}^{(i)}(\ev_{\max}^{(i)})^\herm(\Wm - \Wm^{(i)})\big]\Big) \\
\mathrm{s.t.} &\quad \mathrm{Tr}\big(\Wm^\herm \Am(\beta)\big) \geq \alpha,\quad \forall \beta \in \hat{\mathcal{B}}, \label{flatgain} \\
&\quad \Wm \succeq 0, \label{psd} \\
&\quad \mathrm{Diag}(\Wm) = \mathbf{1}_{N_p}, \label{unitmodulus}
\end{align}
\end{subequations}
where $\Wm = \wv\wv^\herm$, $\Am = E(\beta) (\av(\beta) \odot \qv)(\av(\beta) \odot \qv)^\herm$, $\Wm^{(i)}$ is the solution at iteration $i$, $\ev_{\max}^{(i)}$ is the eigenvector corresponding to the maximum eigenvalue of $\Wm^{(i)}$, and $\eta^{(i)}$ is a penalty factor. The nuclear norm $\|\cdot\|_*$ and spectral norm $\|\cdot\|_2$ promote rank-one solutions.

The convex problem can be solved iteratively using standard solvers (e.g., CVX \cite{CVX}). Upon convergence, the optimized phase-only vector $\wv_{\rm opt}$ is recovered as the principal eigenvector of the final positive semi-definite matrix $\Wm^{\rm opt}$ obtained at the last iteration. 


\bibliographystyle{IEEEtran} 
\bibliography{P2-bibliography}

\end{document}